\documentclass[12pt,draftclsnofoot,onecolumn]{IEEEtran}

\usepackage{cite}
\usepackage{amsmath,amssymb,amsfonts}
\usepackage{algorithm}
\usepackage{algorithmic}
\usepackage{caption}
\usepackage{epsfig}
\usepackage{epstopdf}
\usepackage{graphicx}
\usepackage{textcomp}
\usepackage{xcolor}
\usepackage{booktabs} 
\usepackage{subfigure}
\usepackage{multirow}
\usepackage{rotating}
\usepackage{tikz}
\usepackage{algorithm}
\usepackage{algorithmic}
\usepackage{soul,color}
\usepackage{ifmtarg}
\usepackage{multirow}
\usepackage{microtype}
\def\BibTeX{{\rm B\kern-.05em{\sc i\kern-.025em b}\kern-.08em
    T\kern-.1667em\lower.7ex\hbox{E}\kern-.125emX}}

\begin{document}
\title{MirrorVLC: Optimal Mirror Placement for Multi-Element VLC Networks}
\author{\IEEEauthorblockN{Sifat Ibne Mushfique, Ahmad Alsharoa, \textit{Member, IEEE}, and Murat Yuksel, \textit{Senior Member, IEEE}}\\
\thanks{
Sifat Ibne Mushfique and Murat Yuksel are with University of Central Florida (UCF), Orlando, Florida, USA. Email: sifat.im@knights.ucf.edu, murat.yuksel@ucf.edu.

Ahmad Alsharoa is with Missouri University of Science and Technology (MST), Rolla, Missouri, USA. Email: aalsharoa@mst.edu.}\vspace{-0.5cm}}

\maketitle
\pagestyle{plain}

\begin{abstract}
Visible Light Communication (VLC) is a rapidly growing technology which can supplement the current radio frequency (RF) based wireless communication systems. VLC can play a huge part in solving the ever-increasing problem of spectrum scarcity because of the  growing availability of Light Emitting Diodes (LEDs). One of the biggest advantages of VLC over other communication systems is that it can provide illumination and data communication simultaneously without needing any extra deployment. Although it is essential to provide data rate at a blazing speed to all the users nowadays, maintaining a satisfactory level in the distribution of lighting is also important. In this paper, we present a novel approach of using mirrors to enhance the illumination uniformity and throughput of an indoor multi-element VLC system architecture. In this approach, we improve the Signal-to-Interference plus Noise Ratio (SINR) of the system and overall illumination uniformity of the room by redirecting the reflected LED beams on the walls to darker spots with the use of mirrors. We formulate a joint optimization problem focusing on maximization of the SINR while maintaining a reasonable illumination uniformity across the room. We propose a two-stage solution of the optimization problem with optimization of illumination in the first stage and SINR at the second stage. We propose three different heuristic solutions for the second stage and analyze the performance of them, which demonstrates the advantages of each heuristic for different possible scenarios. We also show that about threefold increase in average illumination and fourfold increase in average throughput can be achieved when the mirror placement is applied which is a significant performance
improvement.
\end{abstract}


\section{Introduction}
Visible Light Communications (VLC) is a rapidly unfolding technology with noteworthy potential to meet the necessity of wireless access speeds. With the constantly increasing number of Internet-of-Things (IoT) devices and the humongous demand for wireless bandwidth, VLC solutions are of lofty worth.
Most of the related work in VLC have concentrated on diffuse optics \cite{2011-wu-power} and diversity combining \cite{2014-Tsiatmas-Optimum} for downloading a data stream to devices in a room. At the modulation level, OFDM \cite{2014-ijaz-Optical, aziz2019high} and Multi-Input-Multi-Output (MIMO) \cite{2014-Nuwanpriya-Angle, 2013-azhar-Gigabit} techniques were explored to increase the VLC link capacities. Unlike diffuse optics, Multi-element VLC (where multiple transmitters are involved in a VLC system design) allows several LED transmitters with narrow divergence angles to transmit different datastreams  simultaneously. Because of this capability, optical wireless communications researchers gained interest in multi-element VLC architectures ~\cite{mushfique2018software,mossaad2015visible,sifat} in the recent times. The multi-element VLC networks can offer increased aggregate throughput via simultaneous wireless links and attain higher spatial reuse. The downlink data transmission efficiency may be considerably enhanced by using multi-element VLC modules due to its light beam directionality where each transmitter, e.g., a Light Emitting Diode (LED), can be modulated with different data streams. Although, with this, the problem of maintaining a balanced illumination also arises because of the potential creation of dark spots in between the directional beams of the transmitters. 

Most of the VLC literature can be categorized into four groups based on the number of transmitters and datastreams involved in the design. \textit{Single Element Single Datastream (SESD)} designs are based on the concept of a single LED transferring a single  datastream to a particular receiver. Most of the studies conducted in the early stages of VLC research were based on SESD VLC systems with a focus on improving the SINR \cite{wang2012performance,liu2014cellular}. \textit{Single Element Multi Datastream (SEMD)} is another form of VLC design where considerable amount of exploration has been done where a single transmitter is able to serve multiple receivers simultaneously or via time sharing methods like Time Division Multiple Access (TDMA) \cite{mmbaga2016performance,abdelhady2017resource} - most of these techniques involved diffuse optics. Apart from these two designs, other two can be categorized in multi-element VLC. \textit{Multi Element Single Datastream (MESD)} design allows multiple LEDs to send data towards one receiver which increases the received light intensity at it. Usage of more than one transmitter allows MESD to employ cooperative transmission~\cite{cooperative-little} to reduce Bit Error Rate (BER) and also various MIMO techniques by placing multiple photo-detectors at the receiver end where such demonstrated an improvement in data rate \cite{butala2013svd}. Despite having these benefits, MESD VLC designs cannot afford high directionality when transferring different data streams to different receiver at the same time, which is possible in \textit{Multi Element Multi Datastream (MEMD)}. These designs have surfaced in recent times, and various combinations in LED positioning are possible in these designs due to the flexibility in the number of elements and datastreams. Our proposed design is under this category where the advantage of light beam directionality is utilized in multi-element VLC system. More details on this categorization can be found in one of our earlier works \cite{mushfique2020optimization}.



Illumination uniformity is a critical element to be examined in a multi-element VLC architecture where uniform light distribution is essential to be maintained as each LED's transmit power is being tuned. Otherwise, inconsistent lighting might appear while the transmit powers of LEDs are being tuned for maximal SINR. It is very important to ensure an uniform lighting distribution in the room, while optimizing the LED assignment problem with a source power constraint on each LED. This type of multi-element architecture can improve the system performance as investigated in~\cite{chen2014improving,Combining_Receiver_2014}. Our work in this paper uses a totally different approach from the previous works in the literature, where we explore a novel approach of using mirror placement in multi-element multi-datastream VLC networks with narrow-angle LEDs in order to use the nLoS light beams for improving overall throughput and the evenness of lighting. 


One of the main challenging tasks in  a multi-element VLC architecture is how to assign group of transmitters (i.e., LEDs) to each receiver in the room at the same time, i.e., serving all users in the room simultaneously and taking the illumination requirements into consideration. There have been several studies to find optimum LED arrangements in MEMD VLC systems to obtain different goals like evenly distributed lighting \cite{burton2012study}, reducing SINR fluctuation \cite{wang2012performance}, improving SINR while maintaining a certain illumination requirement \cite{liu2014cellular}, using cooperative beamforming to optimize total system throughput \cite{cen2019libeam} etc. Our idea is vastly different from these works for mainly two reasons - 1) Optimizing MEMD VLC system design with highly directional LEDs, and 2) Employing mirrors in the walls to use the nLoS component of the VLC channel looking to improve both SINR and illumination uniformity. There have not been much research on using mirrors in optimizing VLC networks. 
In~\cite{park2017novel}, deployment of a double-sized mirror between the receiver photo-detectors is utilized to propose a mirror diversity receiver design which can help in reduction of the channel correlation by attaining two things - obstructing the reception of the light from a particular direction and enhancing the channel gain from another direction. The authors claim this mirror diversity receiver to be an encouraging approach to boost the performance of the VLC system, but the use and benefit from the mirror placement approach are still very limited unlike our work - where we have explored optimum mirror placement in all walls of the room, facilitating to broaden the scope of improvement for both lighting distribution and the SINR.

 In this paper, we propose a novel VLC framework  design with a hemispherical shaped multi-element bulb containing multiple LEDs and several mirrors mounted in the walls. This hemispherical shape allows high spatial reuse in addition to improving the illumination uniformity of the room. Light is emitted in different directions from the LEDs, and hence, the spherical multi-element bulb helps attaining an evenly scattered lighting. We further improve the average SINR and illumination uniformity of the system by formulating and solving an optimization problem, which optimizes the number of the mirrors as well as source power and assignment of the LEDs to the users to improve system performance. Apart from the the multi-element hemispherical bulb, this paper provides a framework using mirror placement for joint optimization of the LEDs' power and association to receivers to improve the aggregate data rate (of multiple data streams) and illumination uniformity in multi-element VLC networks for the first time to the best of authors' knowledge. The main contributions are as follows:

\begin{itemize}
	\item Investigating a hemispherical multi-element bulb architecture in downlink VLC transmission, where each LED can be assigned to a receiver to contribute towards data transmission or towards uniform illumination.
	\item Introducing a mirror placement approach that maximizes the ratio of SINR over the illumination uniformity of the system by formulating it as an optimization problem.
	\item Proposing a two-stage solution of the optimization problem to avoid the impracticality of changing mirror locations based on changes in the position of mobile users.
	\item Proposing multiple ways to place mirrors across the room walls, comparing and analyzing their performance with respect to the case of without any mirror placement.
	\item Proposing multiple low complexity heuristic algorithms to achieve efficient solutions for the second stage of the problem which focuses on maximizing the throughput of the system.
\end{itemize}

The rest of the paper is organized as follows:
Section~\ref{SystemModel} describes our spherical multi-element VLC architecture and the overall system model. The formulation of the proposed joint optimization problem along with the details of the two-stage solution is presented in Section \ref{ProblemFormulation}. Selected numerical simulation results are presented in Section \ref{sec:results}. Finally, we summerize our work in Section \ref{sec:summary}.

\section{System Model}
\label{SystemModel}
				
\begin{figure}[t]
	\centerline{\includegraphics[width=0.95\columnwidth]{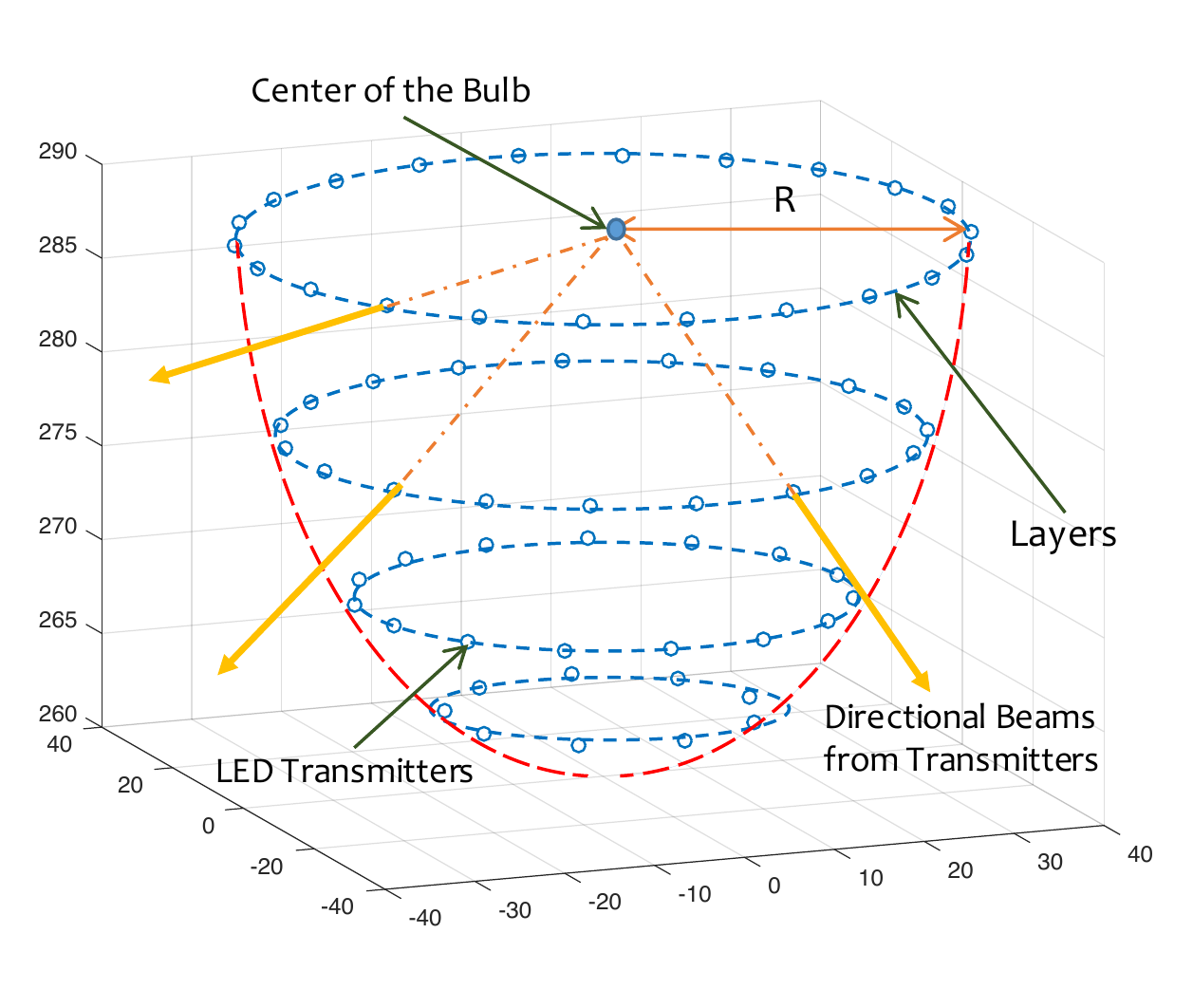}}
	\caption{Placement of transmitters in the multi-element bulb with 4 layers.}\label{Layering}
\end{figure}

We consider a VLC downlink system model in an indoor setting consisting of a single hemispherical bulb containing $M$ directional LED transmitters in $L$ layers to serve $U$ mobile users. The transmit power of each LED is $P_m$ Watt, $\forall m=1,..,M$ as shown in Fig.~\ref{Layering}.

\begin{figure}[t]
	\centerline{\includegraphics[width=0.95\columnwidth]{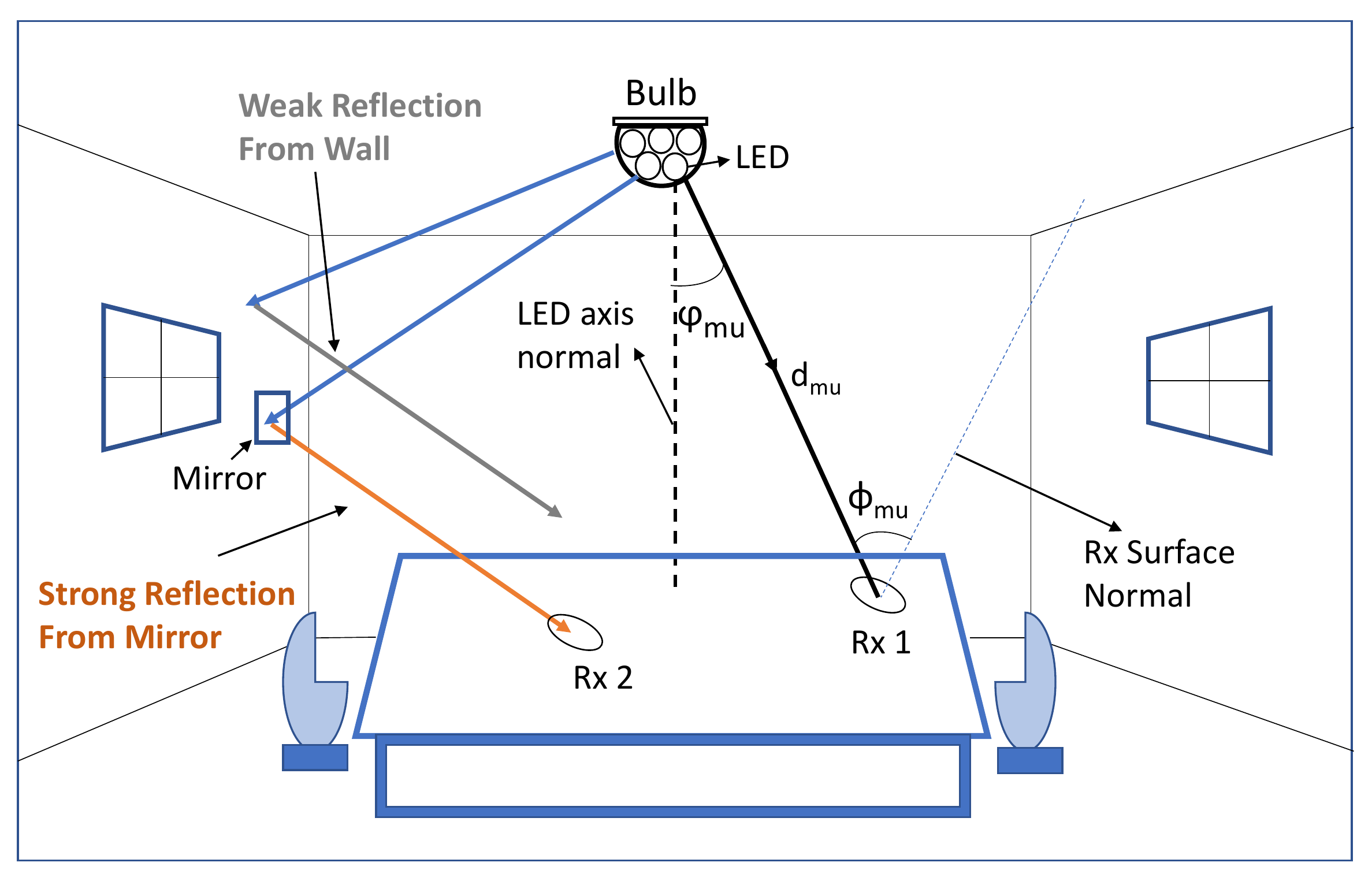}}
	\caption{System Model.}\label{ChannelwithMirror}
\end{figure}

The bulb is a hemispherical structure with two goals: i) achieve uniform lighting illumination within the room, and ii) provide high speed wireless download to mobile users. We assume that the LEDs in the bulb are placed in different layers in order to cover different locations in the room. 
In order to improve the illumination uniformity and download speed to mobile users, we consider placing mirrors to the walls into the room's wall aiming to utilize NLoS beams from the LEDs. We divide each wall into an $X \times Y$ grid. We assume that in each grid element, one mirror can be placed at most. We call this mirror as a 'grid mirror'.
We anticipate that these grid mirrors in the walls will reflect the NLoS signals and thus potentially enhance the performance of the multi-element VLC network. The optimal placement of the mirrors, among other factors, will depend on the direction of the incident lights from the LEDs to the walls.
We consider that a user's Photo-detector (PD) gain depends on the light intensity from three main directions: 1) the LoS beams directly coming from LEDs, 2) strong reflections of the LoS light beams from the grid mirrors, and 3) weak-reflected light from the regular walls as shown in Fig~\ref{ChannelwithMirror}. Since there are four walls and each one is divided as $X \times Y$ sized grid, we visualize the scope of mirror placement as a $4X \times Y$ grid, where four walls are  assumed to be placed side-by-side. Each cell of the grid can be defined with index $z$ where $z = 1,2,3, ..... Z,$ here $Z=4XY$. 

To compute the illumination uniformity in our simulated environment, we consider $N$ fixed sensing points uniformly distributed inside the room. The light intensity received at these points determine how uniform the lighting is inside the room. Although any of them can be placed at a place of interest, we assume that they are uniformly distributed to the room floor in a lattice placement pattern.

We also consider that each mobile user is equipped with a single PD to receive the downlink light beams.
It is assumed that the location of the mobile users in the room can  be predicted. On the other hand, we assume that the mobile users are equipped with an RF transmitter such as Wi-Fi for uplink transmission. In this work, we focus on downlink transmissions only and assume that the uplink transmission speed is not a bottleneck which is also the case for typical wireless access at an indoor setting. 

\subsection{VLC Channel Model}

\subsubsection{LoS Channel}
The LoS channel model between LED $m$ and receiver node $l$ ($l \in \{u \text{ for users}, \newline 
n \text{ for sensing points} \}$) can be expressed as~\cite{ghassemlooy2019optical}:
	\begin{equation}
	\small
	h_{ml}^{\text{LoS}}=\hspace{-.1cm}\left\{
	\hspace{-.1cm}\begin{array}{ll}
	\frac{A_l}{d^2_{ml}} Q_0(\varphi_{ml}) \cos(\phi_{ml}) \hspace{2.25cm}, &\hspace{-.2cm} 0 \leq \phi_{ml} \leq \phi_c \\
	\hspace{-.0cm} 0 \hspace{5.1cm} , &\hspace{-.2cm} \phi_{ml} \geq \phi_c
	\end{array}
	\right.
	\normalsize
	\end{equation}
where $A_l$ is the receiver node's PD area and $d_{ml}$ is the distance between LED $m$ and node $l$. $\varphi_{ml}$ and $\phi_{ml}$ are the irradiance and incidence angles, respectively, as shown in Fig.~\ref{ChannelwithMirror}. 
$\phi_c$ is the FOV angle of the PD. 
We have assumed that no optical filter is used.
$Q_0(\varphi_{ml})$ is the Lambertian radiant intensity and expressed as
\begin{equation}
Q_0(\varphi_{ml})= \frac{(q+1)}{2\pi} \cos^q(\varphi_{ml}),
\end{equation}

where $q=-\ln(2)/ \ln(\cos(\varphi_{1/2}))$ is the order of Lambertian emission and $\varphi_{1/2}$ is the transmitter semi-angle at half power.

\begin{figure}[t]
	\centerline{\includegraphics[width=0.95\columnwidth]{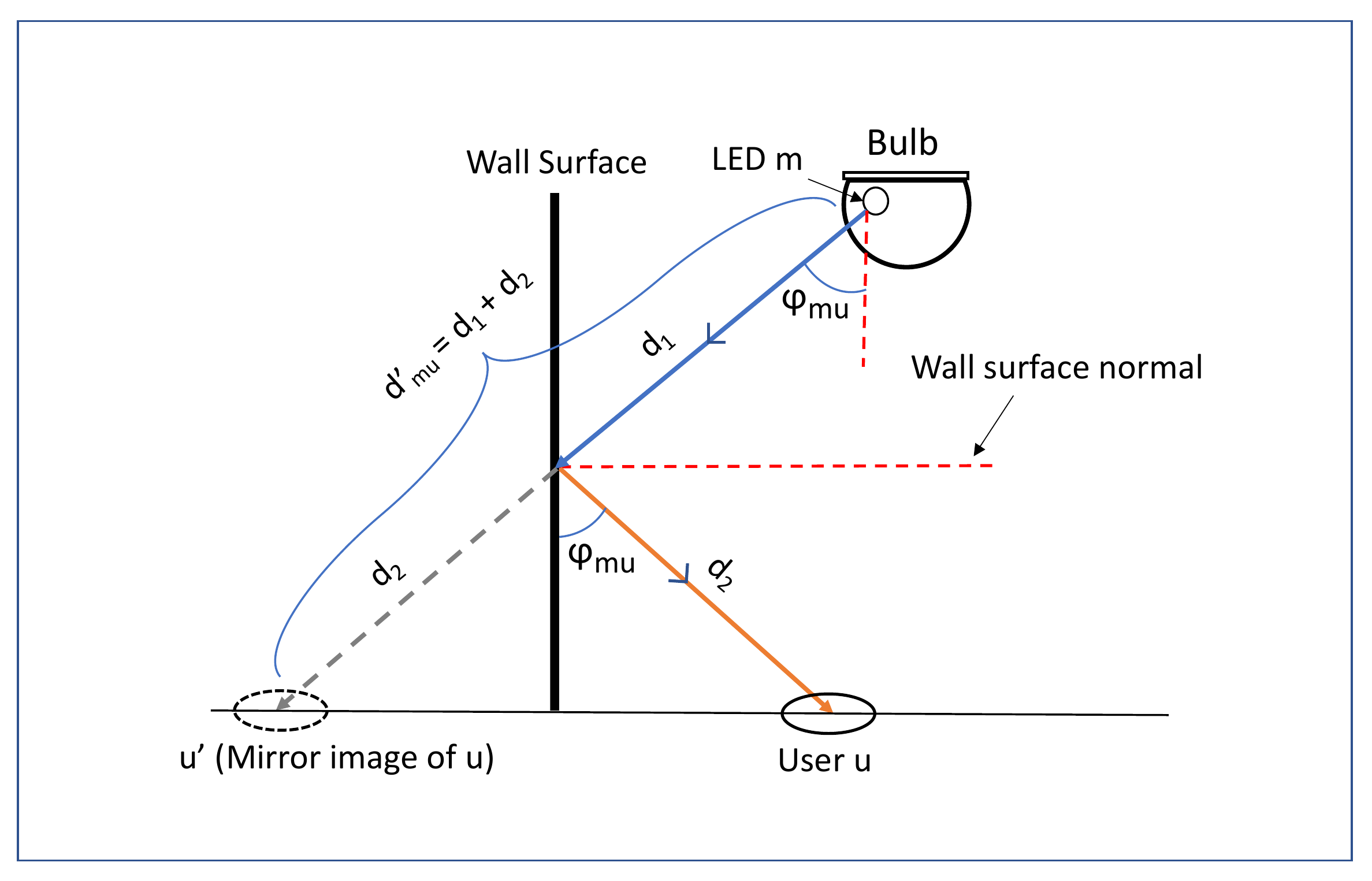}}
	\caption{Comparison between an NLoS channel with mirror and LoS channel by using a mirror image of user $u$ to the wall from the same LED $m$.}\label{NLOS-fig}
\end{figure}

\subsubsection{NLoS Channel}
For simplicity, we consider only the strong reflections of the LoS light beams for computing the NLoS channel model. 
In a scenario where the light from LED $m$ is coming to user $u$ through a grid mirror located at $\xi_{z}$ (i.e., $\xi_{z} = 1$), the NLoS channel model between LED $m$ and node $l$ can be expressed as~\cite{park2017novel}:
\begin{equation}
h_{ml}^{\text{NLoS}}(z)=\frac{\eta A_l}{{\hat{d}_{ml}(z)}^2} Q_0(\varphi_{ml}) \cos(\phi_{ml}) \label{nlos-eqn}
\end{equation}
where $\eta$ is the reflectivity of a grid mirror and $\hat{d}_{ml}(z)$ is the distance between $m$ and $l$ via the mirror located at index $z$ of the grid $\xi$. This is explained in Fig. \ref{NLOS-fig} where we compare the NLoS channel between LED $m$ and user $u$ with a channel between $m$ and $u'$ which is a mirror image of $u$ with respect to the wall. We can see that the distance between $m$ and $u'$ is same as the total distance between $m$ and $u$ via the mirror. Therefore, we can express the NLoS channel between $m$ and $u$  as the LoS channel between $m$ and $u'$ with the following in consideration - the NLoS channel strength will be reduced by a factor as there is a mirror in its path, and this factor is dependent on $\eta$, the reflectivity of the mirror. Same expression can be derived for the sensing points as well, thus we present the general expression for node $l$ in \eqref{nlos-eqn}.

\subsubsection{Total Channel}
We can express the total channel model as the combination of LoS and NLoS channels as below:
\begin{equation}
H_{ml} = h_{ml}^{\text{LoS}}+ \sum\limits_{z=1}^{Z} \chi_{mz} h_{ml}^{\text{NLoS}}(z)
\label{totalch}
\end{equation}
\hspace{1.9cm} such that
\begin{equation}
\begin{aligned}
\hspace{.5cm}\chi_{mz} = \xi_z, & \hspace{2cm}  if \hspace{2mm} z \in {\Upsilon}_{zm}, \\
\hspace{-.1cm}\chi_{mz} = 0, & \hspace{2cm} otherwise.
\end{aligned}
\label{eqn:grid0}
\end{equation}

To explain the above channel clearly, we define \emph{reflection area} of LED $m$ as the wall area being covered by the beam coming from LED $m$.

\begin{figure}[t]
	\centering
	{\includegraphics[width=0.95\columnwidth]{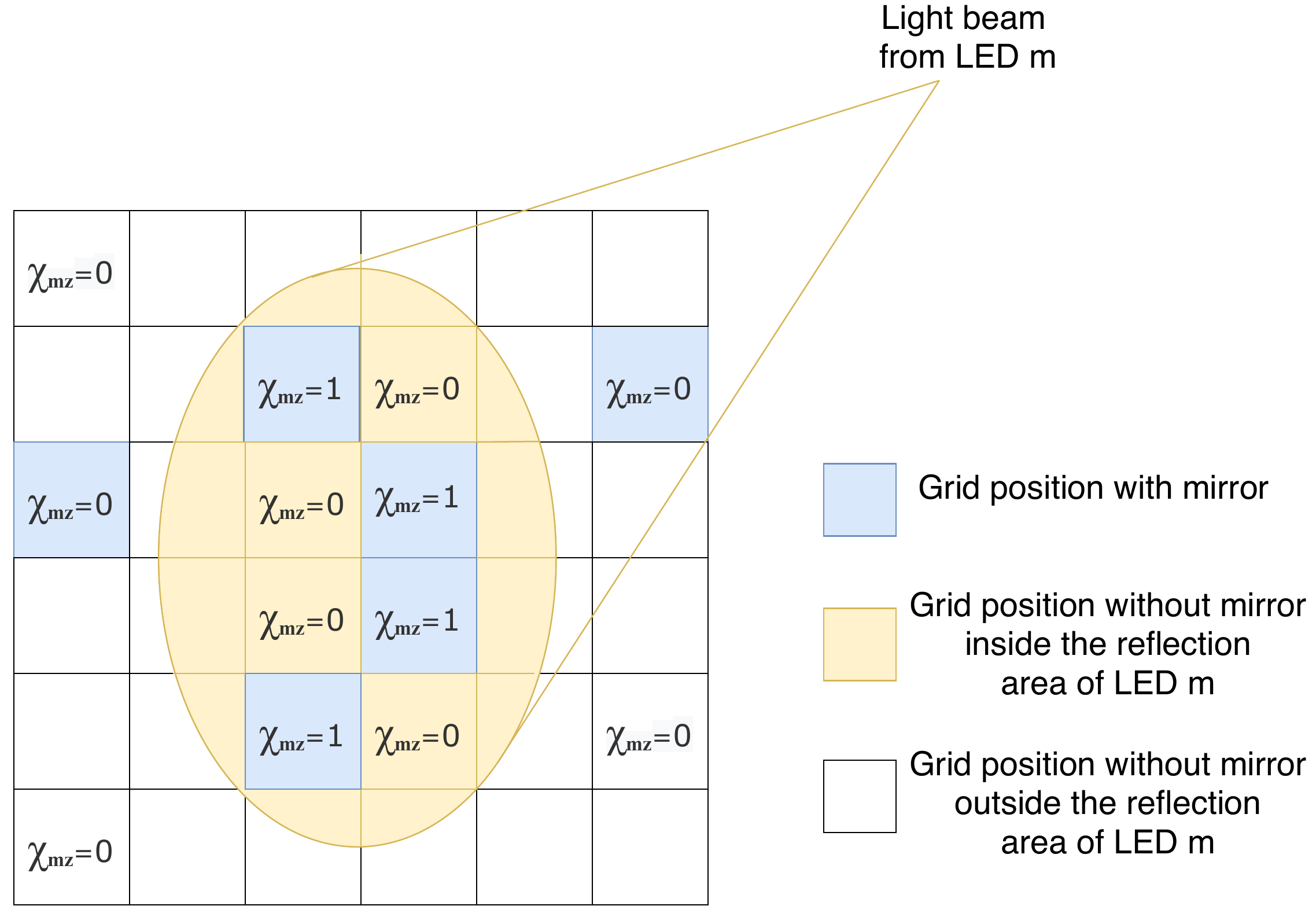}
	}
	\caption {Figure explaining different values of $\chi_{mz}$ in a room wall. $\chi_{mz} = \xi_z$ when grid position $z$ is inside the reflection area of LED $m$ (marked in orange) and 0 otherwise. Grid positions with mirrors are marked in light blue.}
	\label{fig:chi}
\end{figure}

$\Upsilon_{zm}$ in eq. \ref{eqn:grid0} is the set of $z$ indices corresponding to the reflection area of LED m. $\xi_z = 1$ if a grid mirror is placed at the cell located at index $z$ of the grid $\xi$ and $\xi_z = 0$ otherwise. 
Further, $\chi_{mz}$ is a binary variable that expresses whether or not a NLoS component from LED m via grid location $z$ will be accumulated in the channel. Note that $\chi_{mz}$ cannot be 1 if the grid location $z$ is not within the reflection area of LED $m$. However, if the grid location $z$ is within the reflection area of LED $m$, then it can be 0 or 1 depending on the existence of a grid mirror at $z$ which is expressed by $\xi_z$. 
Fig. \ref{fig:chi} explains different values for $\chi_{mz}$.

Although the above explained model seems a bit complex primarily because of the inclusion of an extra dependent variable $\chi_{mz}$, a designer of a VLC system can enjoy much more flexibility because of this. By bringing  different combinations of $\chi_{mz}$ into the overall search space, probability of finding a near optimal solution becomes much higher as we can control the contribution of every small fractions of each LED beam to either illumination only, or both illumination and communication.

\subsection{Illumination Uniformity}
The illumination intensity distribution across the room is an important factor to be considered in VLC system. One of common metric to measure the illumination intensity is the illumination uniformity. The illumination uniformity, $\vartheta$, can be defined as the ratio between the minimum and the average illumination intensity among all
N sensors and is given as~\cite{standardization2002lighting}:
\begin{equation}
\label{eqn:uni-exp}
\vartheta=\frac{\underset{n}\min\left(\sum\limits_{m=1}^M \alpha_0 P_{m} H_{mn}\right)}{\frac{1}{N}\sum\limits_{n=1}^N  \sum\limits_{m=1}^{M} \alpha_0 P_{m} H_{mn}}
\end{equation}
where $\alpha_0$ is the luminous efficiency that depends on the LED color wavelength, e.g. $\alpha_0=60$ lumen/watt for white LED~\cite{efficacywiki}. $\min(.)$ is the minimum function.

\subsection{LED-User Association}
We use a binary variable $\epsilon$ that indicates the association between LED $m$ and user $u$ which is given as follows:
\begin{equation}
\epsilon_{mu}=\hspace{-.1cm}\left\{
\begin{array}{ll}
\hspace{-.2cm}1, \hspace{-.3cm}& \hbox{if LED $m$ is associated with user $u$.} \\
\hspace{-.2cm}0, \hspace{-.3cm}& \hbox{otherwise.}
\end{array}
\right.
\end{equation}
We assume that user $u$ can be associated with multiple LEDs at the same time. However, it is assumed that each LED can be associated with one user at most during the same time. Therefore, the following conditions should be respected: 	
\begin{equation}
\sum\limits_{u=1}^U \epsilon_{mu} \leq 1, \forall m=1,..,M.
\label {eqn:eps1}
\end{equation}

\subsection{SINR Calculation}
We assume that each LED is either associated with one user or used for lighting only. Therefore, SINR at user $u$ can be expressed as ~\cite{8113471}
\begin{equation} \label{eqn:sinr-exp}
	\Gamma_u=\frac{ \left(\sum\limits_{m=1}^{M} \epsilon_{mu} H_{mu} P_{m} \right)^2}{N_0 B+ \sum\limits_{\substack{k=1 \\ k\neq u}}^U \left(\sum\limits_{m=1}^{M} \epsilon_{mk} H_{mk} P_{m}  \right)^2 }
\end{equation}

where $B$ and $N_0$ are the communication bandwidth and the spectral density of the Additive White Gaussian Noise (AWGN), respectively.


\section{Problem Formulation and Solution}\label{ProblemFormulation}
In a nutshell, we propose an optimization problem to optimize the combination of SINR and illumination uniformity based on mirror placement on the wall ($\xi_z$), LED-user assignment ($\epsilon_{mu}$) and source power for each of the $m$ LEDs ($P_m$). Although it is possible to re-optimize the objective value by updating $\epsilon_{mu}$ whenever a user changes its position, it is not feasible to do so for the mirrors. For this reason, we have divided the problem in two separate optimization problems. First, we find the optimal mirror placements and LED source powers that maximize the illumination uniformity, and then, we optimize the  LED-user association along with tuning the LED source powers to maximize the combination of SINR and illumination uniformity using the results from the first problem. This staging makes the optimal mirror placement more practical. In particular, the solution to the first problem will yield the best places to set up the grid mirrors so that high illumination uniformity is attained. Then, the second problem can be solved on-the-fly as users are moving in the room, yielding the best LED transmit powers and LED-user association depending on the user movements.

In this section we formulate two optimization problems defined as mirror design problem and communication problem. 

\subsection{Design Problem}\label{design}

The main goal of this problem is to achieve the best illumination quality by optimizing not only the mirror placements but also the LEDs' transmit powers. Therefore, the mirror design optimization can be formulated as follows:
\begin{align}
	&\hspace{-0.5cm}\underset{\substack{\xi_ z, \chi_{mz} \in\{0,1\}, P_m}}{\text{maximize}} \quad  \underset{n}\min\left(\sum\limits_{m=1}^M \alpha_0 P_{m} H_{mn}\right)  \label{of_design}\\
	&\hspace{-0.5cm}\text{subject to \eqref{eqn:grid0} and:}\nonumber\\
	&\hspace{-0.5cm} P_{min} \leq P_m \leq \bar{P},  \quad \forall m,\label{power0}\\
	&\hspace{-0.5cm} 
	\frac{\underset{n}\min\left(\sum\limits_{m=1}^M \alpha_0 P_{m} H_{mn}\right)}{\frac{1}{N}\sum\limits_{n=1}^N  \sum\limits_{m=1}^{M} \alpha_0 P_{m} H_{mn}} \geq \mu \label{unifor_constraint0}
	\\
	&\hspace{-0.5cm} \phi_2 \leq \sum\limits_{m=1}^M \alpha_0 P_{m} H_{mn} \leq \phi_1, \quad \forall n,  \label{minimum_illumination}
\end{align}
where $P_{min}$ and $\bar{P}$ are the minimum and maximum source power respectively that can be allotted to an LED and $\mu$ is the minimum acceptable illumination uniformity for an indoor setting~\cite{standardization2002lighting}. $\chi_{mz}$ is a dependent variable here, i.e., it is directly related to the values of $\xi_z$. $\phi_1$ and $\phi_2$ are the maximum and minimum levels of total illumination allowed at a particular sensing point. These are introduced to ensure that the room is illuminated at a minimum level and also it does not go beyond a level where it can be harmful to human eye. 

We aim to solve the above design optimization problem optimally.
In order to do this, we first linearize the objective function \eqref{of_design} and constraint \eqref{minimum_illumination} by introducing a new decision variable $\phi$ as follows:
\begin{equation}\label{SNR}
\phi= \underset{n}\min\left(\sum\limits_{m=1}^M \alpha_0 P_{m} (h_{mn}^{\text{LoS}}+ \sum\limits_{z=1}^{Z} \chi_{mz} h_{mn}^{\text{NLoS}}(z))\right).
\end{equation}
Equation~\eqref{SNR} can be re-written as follows:
\begin{equation}\label{SNR2}
	\phi= \underset{n}\min\left(\sum\limits_{m=1}^M \alpha_0 P_{m} h_{mn}^{\text{LoS}}+ \sum\limits_{z=1}^{Z} \alpha_0 \rho_{mz} h_{mn}^{\text{NLoS}}(z)\right),
\end{equation}

\noindent where $\rho_{mz}=\chi_{mz} P_m$, which is introduced as another decision variable to linearize the product of binary variable $\chi_{mz}$ and real decision variables $P_m$.
Indeed maximizing the objective function given in \eqref{of_design} is equivalent to maximizing $\phi$ given that $\phi \leq \sum\limits_{m=1}^M \alpha_0 P_{m} H_{mn})$, $\forall n$.
With the introduction of $\rho_{mz}$, the following inequalities also have to be respected:
\begin{align}
&1) \quad P_m \geq \rho_{mz} \geq 0, \forall m, \forall z \notag\\
&2) \quad \rho_{mz} \geq \bar{P}_m \chi_{mz} - \bar{P}_m + P_m, \forall m, \forall z \notag\\
&3) \quad \rho_{mz} \leq \bar{P}_m \chi_{mz}, \forall m, \forall z.~\label{Linear}
\end{align}

The first two inequalities ensure that $\rho_{mz}$ value is between $\chi_{mz}$ and $P_m$. The third inequality guarantees that $\rho_{mz}=0$ if $\chi_{mz}=0$, and $\rho_{mz}=P_m$ if $\chi_{mz}=1$.
Therefore, the optimization problem can be reformulated as binary linear optimization problem as follows:
\begin{align}
	&\hspace{-0.5cm}\underset{\substack{\xi_ z, \chi_{mz}, P_m \\ \phi, \rho_{mz}}}{\text{maximize}} \quad  \phi  \label{of_design2}\\
	&\hspace{-0.5cm}\text{subject to \eqref{eqn:grid0}, \eqref{power0}, \eqref{Linear}, \eqref{minimum_illumination} and:}\nonumber\\
	&\hspace{-0.5cm} \phi \geq \frac{\mu}{N}\sum\limits_{n=1}^N  \sum\limits_{m=1}^{M} \alpha_0 \left(P_{m} h_{mn}^{\text{LoS}} + \sum\limits_{z=1}^{Z} \rho_{mz} h_{mn}^{\text{NLoS}}(z)\right) \label{unifor_constraintm2}\\
    &\hspace{-0.5cm} \phi \leq \sum\limits_{m=1}^M \alpha_0 \left(P_{m} h_{mn}^{\text{LoS}} + \sum\limits_{z=1}^{Z} \rho_{mz} h_{mn}^{\text{NLoS}}(z) \right), \forall n \label{minimum_illumination3}
\end{align}

It can be noticed that the solution for such binary linear optimization problem can be determined optimally using on-the-shelf software such as Gurobi/CVX interface~\cite{Gurobi}.

\subsection{Communication Problem}\label{communication}

After solving the mirror placement design optimization problem, we focus now on solving the communication phase aiming to maximize the SINR of the system to ensure the best possible data rate to the users. We choose to use Max-Min utility of the SINR. The approach of maximizing the total data rate which is known in the literature as Max C/I \cite{bifairness}, promotes users with favorable channel and interference conditions by allocating them most of the resources, whereas users suffering from higher propagation losses and/or interference levels will have very low data rates. Therefore, due to the unfairness of total sum data rate utility, the need for more fair utility metrics arises.  The Max-Min utilities are a family of utility functions attempting to maximize the minimum SINR in a network~\cite{Min-Max}. Our goal is to induce more fairness in the network by increasing the priority of users having lower SINR using the Max-Min utility. Thus, we formulate an optimization problem aiming to maximize the minimum SINR of all users by taking the association and illumination intensity constraints into consideration. This optimization problem can be expressed as	
 	
\begin{align}
&\hspace{-0.5cm}\underset{\substack{\epsilon_{mu} \in\{0,1\}, P_m \geq 0}}{\text{maximize}} \quad  \Gamma_{min} \label{of1}\\
&\hspace{-0.5cm}\text{subject to \eqref{eqn:eps1}, \eqref{power0} and \eqref{unifor_constraint0},}\nonumber
\end{align}	

\noindent where $\Gamma_{min} = \underset{u}\min(\Gamma_u)$ is the minimum SINR among all users.
\\

\subsubsection{NP-Completeness}
As the number of LEDs $M$ being considered can be quite large, the number of ways to assign these LEDs to multiple users can be very large and it is practically infeasible to obtain an optimal solution as the users might move frequently inside the room, and the problem is needed to be solved again whenever any of the user coordinates is updated. In fact, this LED assignment problem in consideration is proven to be NP-Complete as elaborated in one of our earlier works~\cite{mushfique2020optimization}, so we propose three low complexity heuristic solutions to solve it. 
\\
\subsubsection{Heuristic Approaches}
We describe each of our proposed heuristic approaches in this subsection. 
\begin{figure}[t]
	\centering
	{\includegraphics[width=0.85\columnwidth]{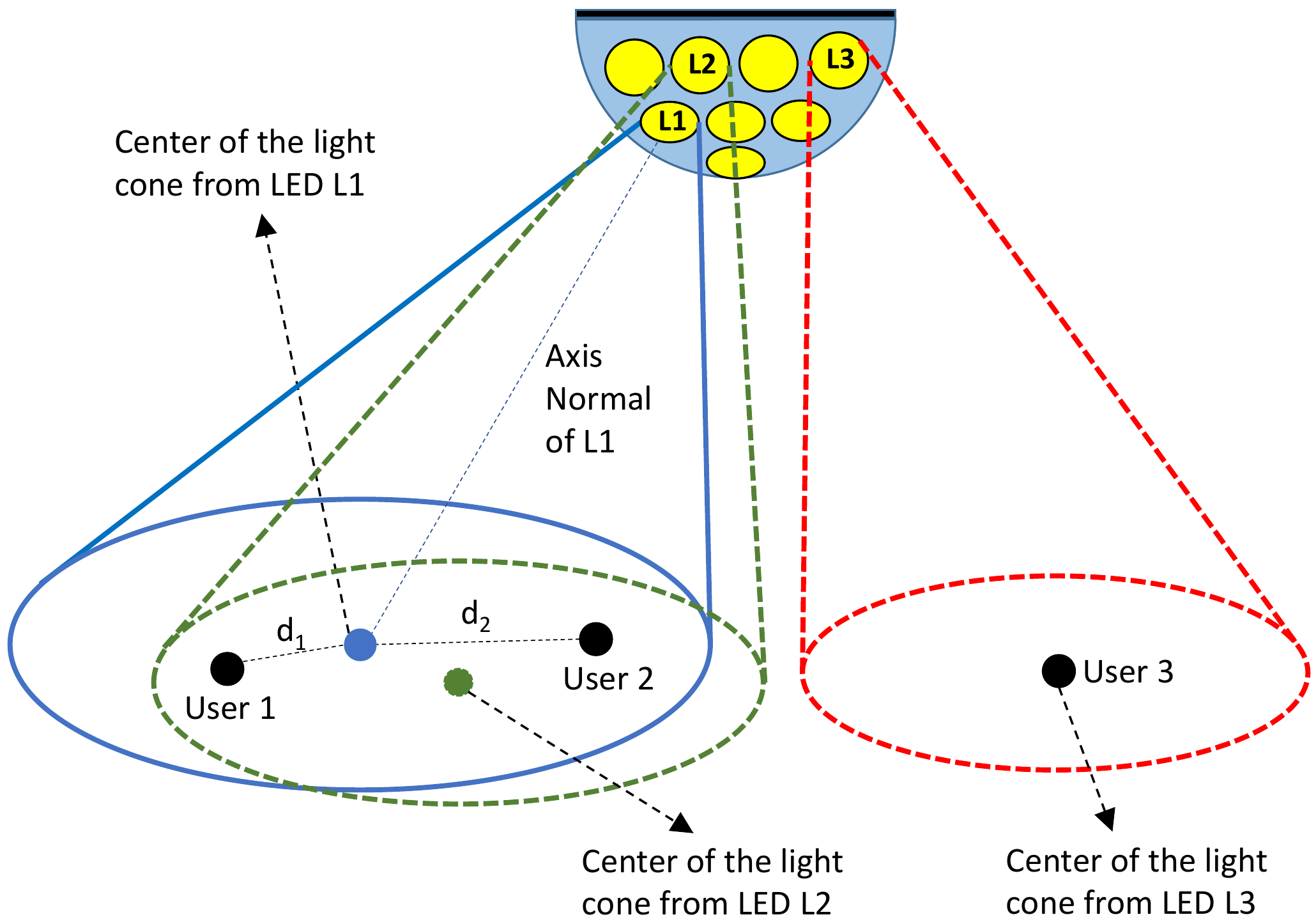}
	}
	\caption {A case of interference for an user-by-user approach such as \textit{NUA} or \textit{SSA-LED}: LEDs L1, L2 and L3 are assigned to user 1, 2 and 3 respectively. In this case, received power from L2 and L3 will be considered as interference at user 1 - since these LEDs are assigned to user 2 and user 3.}
	\label{fig:cone}
\end{figure}

\vspace{1mm}\noindent\emph{Nearest User Assignment (NUA)}: 
We denote the resulting $P_m$ values from the solution of \eqref{of_design2} as intermediate power values $P^{prev}_{1}$, $P^{prev}_{2}$, .... $P^{prev}_{M}$ for $M$ LEDs, and  assign LED $m$ to the closest user to its beam projection if the user lies in the cone of LED $m$.
If there is one or no user in the cone, then we allocate $\min(P^{prev}_{m}(1+\tau),\bar{P})$ where $\tau$ is a measurement of how much we can deviate the allocated power to LED $m$ from its intermediate value $P^{prev}_{m}$. We use 0.1 as the value for $\tau$ in our simulations.
If there is more than one user in the cone of LED $m$, then we assign this LED to the user which is nearest to the center of the cone of the LED, but with fractional power, since there will be interference in this case as shown in Fig. \ref{fig:cone}. More details on this heuristic algorithm can be found in \cite{mushfique2020optimization} where it was introduced for the first time.

\begin{figure}[t]
	\centering
	{\includegraphics[width=0.85\columnwidth]{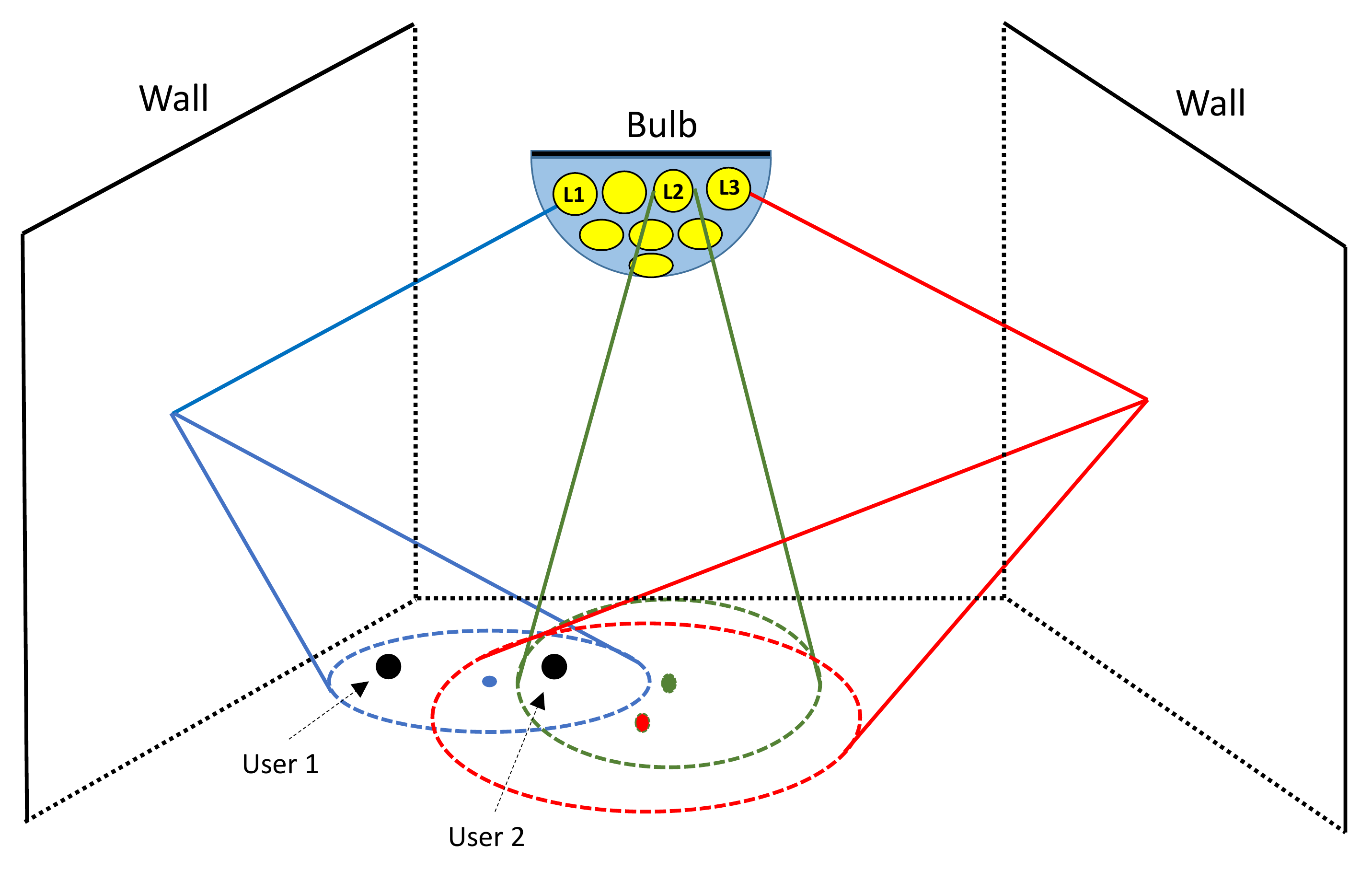}
	}
	\caption {A case of conflict for an user-by-user approach such as \textit{SSA-User}: both User 1 and User 2 are getting strongest signal from LED L1, but since L1 is assigned to User 1 first, User 2 has to be assigned to L2, which provides the second strongest signal for User 2.}
	\label{fig:conessa}
\end{figure}

 	\vspace{1mm}\noindent\emph{Strongest Signal-based Assignment with User-first Approach (SSA-User)}:
	An LED-by-LED assignment approach is implemented in the \textit{NUA} algorithm. In contrast, in this approach, we do a user-by-user assignment. We inspect each user one by one and check the user in consideration is under how many LED beams, either via LoS or via NLoS. There are three possible scenarios - 
	
	1) If there is no incoming LED beam towards this user, then we do not assign any LED to it. 
	
	2) If there is only one incoming LED beam, then we follow similar approach to \textit{NUA} and assign the user with source power $\min(P^{prev}_{m}(1+\tau),\bar{P})$. 
	
	3) If there are more than one incoming LED beams, then we take total channel model values  between each of these LEDs and this particular user into consideration. There are two possible cases in this scenario - i) There is no incoming LED  which is already assigned to another user, which means no possible interference - and ii) There is one or more LED(s) which are already assigned to other user(s), which creates interference. 
	In the first case, since there is no chance of interference from any other user, we assign all the incoming LEDs to the user in consideration. However, a much more careful approach is needed for the second case, which is shown in Fig. \ref{fig:conessa}. Here we assign all the incoming LED beams, or simply `incoming LEDs', to the current user which are not already assigned to other user(s). 
 We calculate the fraction, $\kappa$, as the ratio of maximum channel value of all the incoming LEDs including those also which are already assigned to other user(s) denoted as $H_{m_ju_i}$ assuming $m_j$ is the one among these LEDs with strongest channel value with the current user $u_i$ - and maximum of the channel values between all the incoming LEDs to the current user which are still unassigned, denoted as $H_{m_iu_i}$ assuming $m_i$ is the one among these LEDs with strongest channel value with $u_i$. It can be expressed as $\kappa = \frac{H_{m_ju_i}}{H_{m_iu_i}}$. Finally, we allot the LED $m_i$ with the source power of $\max ( P_m^{prev}(1-\kappa), P_m^{prev}((1-\tau))$.	 
	
	\vspace{1mm}\noindent\emph{Strongest Signal-based Assignment with LED-first Approach (SSA-LED)}: This approach is a blend of the previous two approaches - the scanning procedure is LED-by-LED like \textit{NUA} and the assignment is based on channel strength like \textit{SSA-User}. For each LED $m$ in the bulb, this approach checks whether there is any user inside it's coverage, either directly or via the mirror. We allocate $\min(P^{prev}_{m}(1+\tau),\bar{P})$ amount of source power when there is no or one user is covered by this LED. Now, if there are more than one user covered, we assign the current LED $m$ to the user $u_i$ with the strongest channel value only if the number of the users covered by this LED is less than or equal to 3 per cent of the total number of users. This is done to reduce the effect for interference when multiple users are being covered by the same LED. We give fraction of $P^{prev}_{m}$ to the current LED with a ratio $\kappa = \frac{H_{mu_j}}{H_{mu_i}}$, where $u_j$ is the user with the second strongest channel value with the current LED $m$. Thus the power allocated to LED m is $\max ( P_m^{prev}(1-\kappa), P_m^{prev}((1-\tau))$.	 
	
	\vspace{1mm}
	Each of the above described approaches has it's own advantages and disadvantages. \textit{NUA} has the simplest approach, although it is the worst-performing one which is elaborated further in Section \ref{sec:results}. \textit{SSA-User} is suitable for the scenarios where the users in the room are prioritized in some manner as it is an user-by-user approach which provides the best service to the user which is chosen first. But this is the most expensive approach considering the time complexity as many of the LEDs are needed to be examined multiple times if they cover more than one user. \textit{SSA-LED} can be considered as a balance between \textit{NUA} and \textit{SSA-User} since it is less complex than \textit{SSA-User} and obtains better performance than \textit{NUA}. It also provides fairness to all the users in the system.

\subsubsection{Computational Complexity}
We are assuming $M$ total LEDs and $U$ total users in our problem. In both \textit{NUA} and \textit{SSA-LED}, we check each of the LEDs to allocate source power and assign it to a user if needed. This effective runtime of checking this is $O(M)$. Calculating the channel values for all users will take $O(MU)$ time which is the most dominant component, so the overall runtime for both \textit{NUA} and \textit{SSA-LED} is $O(MU)$.

For \textit{SSA-User}, we take an user-by-user approach, and there may be cases when there is a conflict while assigning an LED to a particular user. In the worst case, runtime of this will be $O(M^2)$ when but on average this will be $O(MlogM)$ as this situation is similar as the Quick Sort algorithm. Combining this with the running time $O(MU)$ to calculate the channel values, we find the overall time complexity of the \textit{SSA-User} approach to be $O(M(U+logM))$.

\section{Simulation Results} \label{sec:results}

 \begin{table}[t]
 	\centering
 	\caption{Simulation Parameters}
 	\label{tab:parameters}
 	\begin{tabular} {	
 			| p{4.0in} | p{1.99in} |}
 		\hline
 		\textbf{Parameter} & \textbf{Value} \\
 		\hline
 		Room Size & $6m \times 6m \times 3m$ \\
 		Radius of the hemispherical bulb, $R$ & $40cm$ \\
 		Number of LEDs on the bulb, $M$ & $391$ \\
 		Radius of an LED transmitter, $r_t$ & $1.5cm$ \\
 		Divergence angle of the LEDs, $\theta_d$ & $30^o$ \\
 		Maximum transmit power of the LEDs, $\bar{P}$ & $0.1W$ \\
 		Area of a PD receiver$^*$, $r_r$ & $5cm^2$ \\
 		Area of a sensing point, $r_s$ & $10cm^2$ \\
 		Number of users, $U$ & $2-12$ \\
 		Number of sensing points, $N$ & $100$ \\
 		Visibility, $V$ & $0.5km$ \\
 		Optical signal wavelength, $\lambda$ & $650nm$ \\
 		AWGN spectral density, $N_0$ & $2.5 \times 10^{-20} W/Hz$ \\
 		Modulation bandwidth, $B$ & $20MHz$ \\
 		Minimum uniformity, $\mu$ & $0.7$ \\
 		Minimum illumination, $\phi_2$ & $400$ $lux$ \\
 		Maximum illumination, $\phi_1$ & $600$ $lux$ \\
 		\hline
 	\end{tabular}
 	\thanks \vspace{3mm} {$^*$We assume that an array of PDs is used to attain a large receiver area.}
 \end{table}

In this section, we provide simulation results to study the performance of our MEMD VLC system model. Our aim is to make comparison between the three heuristics we proposed in terms of the \emph{minimum throughput} and the \emph{average throughput} of the system. To compare our proposed methods (\textit{NUA}, \textit{SSA-LED}, and \textit{SSA-User}), we use the same input parameters for all of them. We randomly place the users on the room floor with their receiver's FOV normal looking towards the ceiling. We report the average of the minimum throughout and illumination uniformity results among these randomly generated cases. To gain confidence in our results, we repeated the simulation experiments 100 times for all the results.

We use white LEDs with luminous efficacy $\alpha_0=169$  lumen/watt~\cite{pattison2018led}. The default values of the remaining input parameters used in our simulations are given in Table~\ref{tab:parameters}. For placing LEDs on the bulb, we followed the method in Section~\ref{SystemModel}. In particular, for a bulb with $R$ = 40cm radius, we place $M$ = 391 LEDs on 20 layers with $k_{1..20}$ = [1, 6, 12, 15, 19, 26, 30, 37, 43, 33, 30, 28, 25, 21, 16, 13, 11, 10, 9, 6].

\begin{figure}[t]
	\centering
	{\includegraphics[width=0.95\columnwidth]{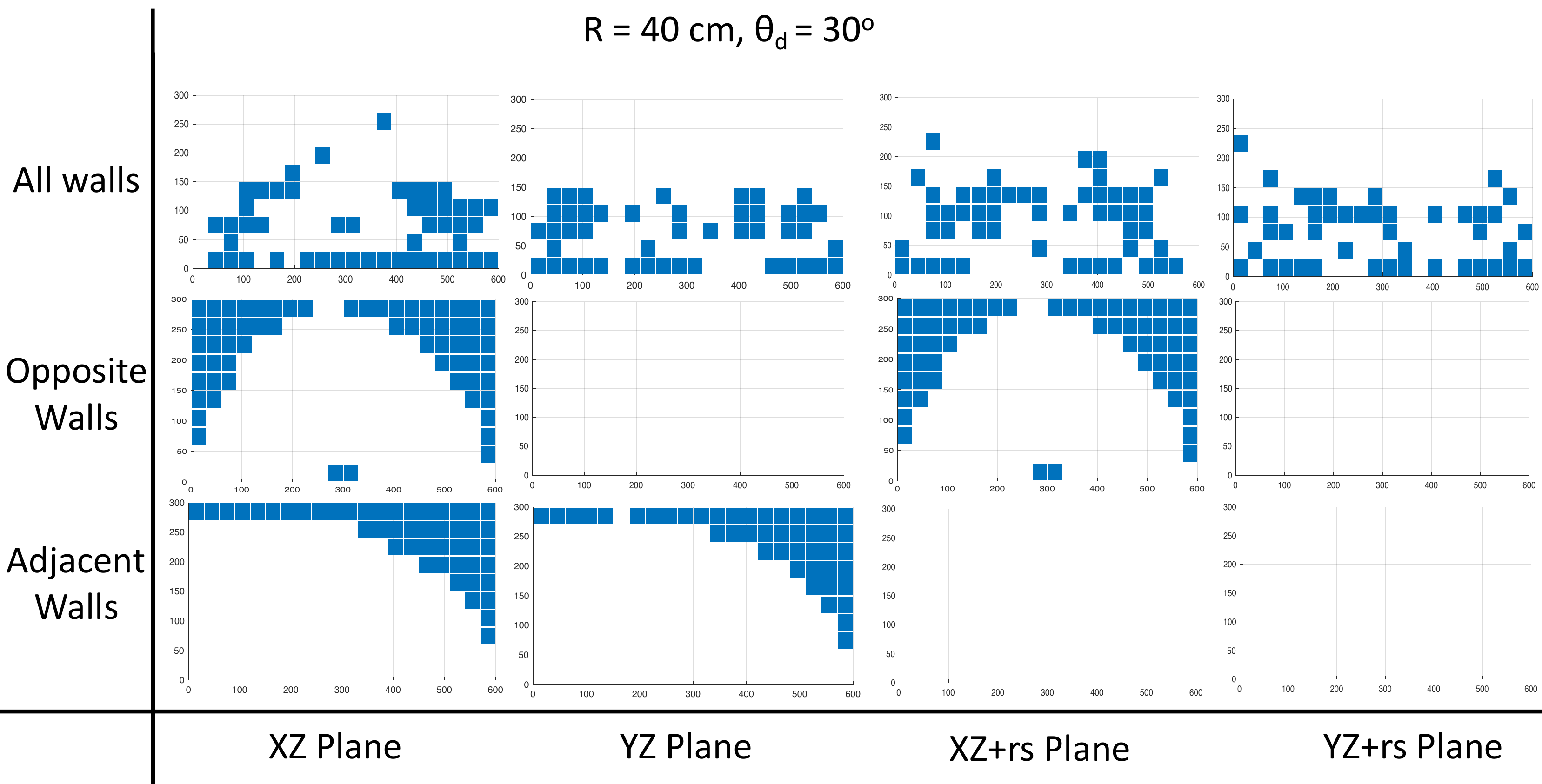}
	}
	\caption {Mirror placement heatmaps in four walls for different bulb radiuses and divergence angles.}
	\label{fig:heatmap}
\end{figure}

\subsection{Placement of Mirrors}
We solve the design problem in \eqref{of_design2} using different mirror placement approaches for different scenarios. For $R = 40 cm$ and  $\theta_d=30^o$, we solve the design problem considering three separate approaches : Mirror placement on 1) two adjacent walls, 2) two opposite walls and 3) all four walls. We define the four walls as $XZ$, $YZ$, $XZ+rs$ (the plane parallel with  $XZ$-plane) and $YZ+rs$ (the plane parallel with $YZ$-plane) planes, where $rs$ can be denoted as the room floor size. The mirror placement heatmaps from each of these scenarios are shown in Fig. \ref{fig:heatmap}. We can see that when we let the optimizer place mirrors on all four walls, it is difficult to establish any concrete pattern from the heatmaps although there are some similarities in placing the mirrors at lower altitudes. This happens because when the optimizer considers the sensing points near the corners of the room, there are several options to choose a mirror location from either of the two adjacent walls sharing the corner, and since there is no particular rule to do that, the mirrors are chosen randomly from those possible locations which could yield similar performance in terms of illumination level and uniformity, thus breaking the symmetry in the heatmaps. This is not the case when we allow the optimizer to place mirrors in two opposite walls as they do not share any corner, so we can see the same pattern of mirror placement in those two walls. Even in the case of two adjacent walls, the mirror placement follows the same pattern as the optimizer can avoid placing the mirror on the sides of the two walls which share the same corner and place most of the mirrors of the sides which do not share the same corner, as illustrated in the figure.

\subsection{Performance Comparison among Mirror Placement Approaches}
\begin{figure*}[t]
 	\centering
 	\subfigure[Minimum throughput vs. No. of users. ]{
 		\includegraphics[width=0.31\columnwidth] {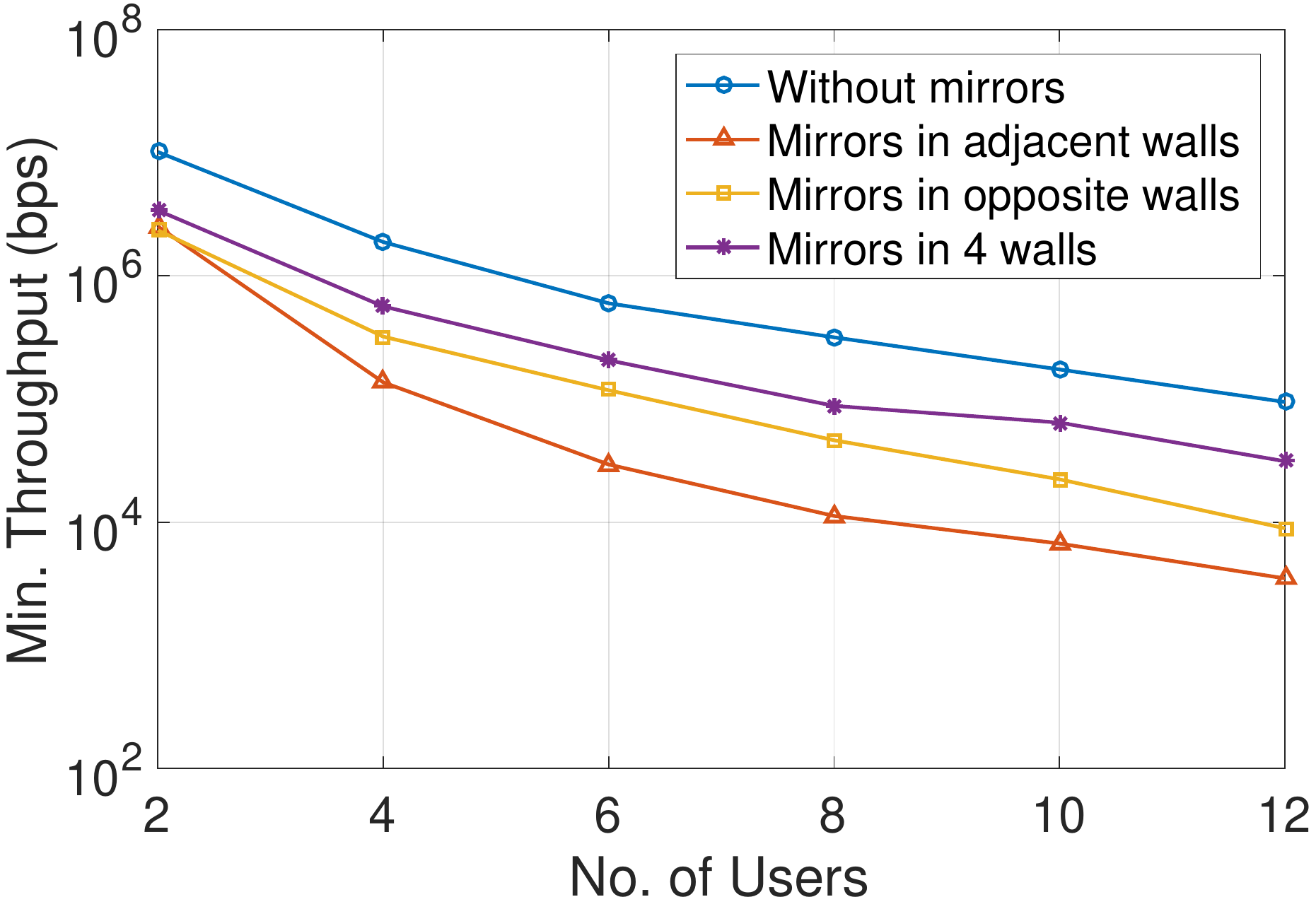}
 		\label{fig:mintp-nua}
 	} 
 	\subfigure[Average throughput vs. No. of users.]{
 		\includegraphics[width=0.31\columnwidth] {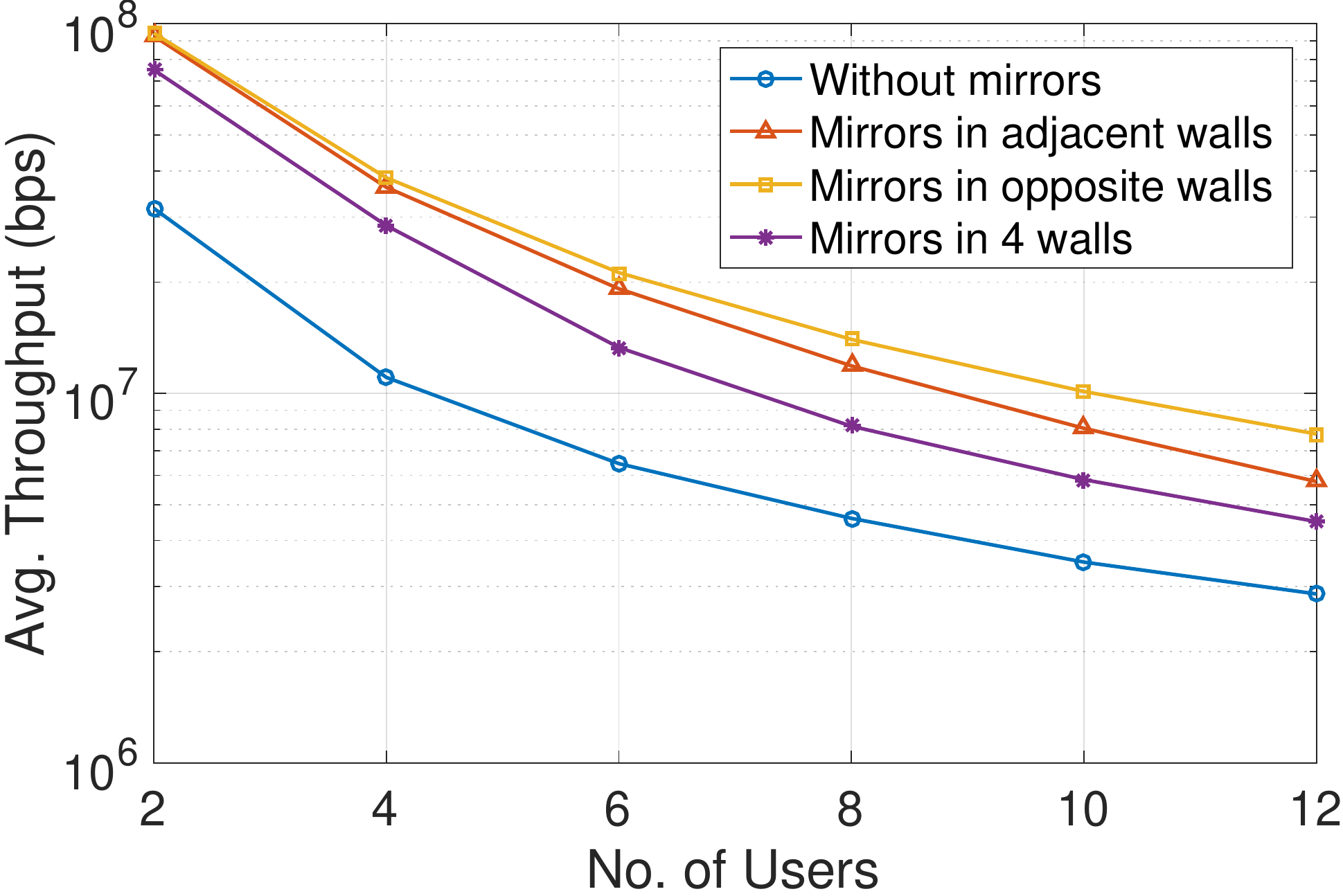}
 		\label{fig:avgtp-nua}
 	} 	
 	\subfigure[Average illumination vs. No. of users.]{
 		\includegraphics[width=0.31\columnwidth] {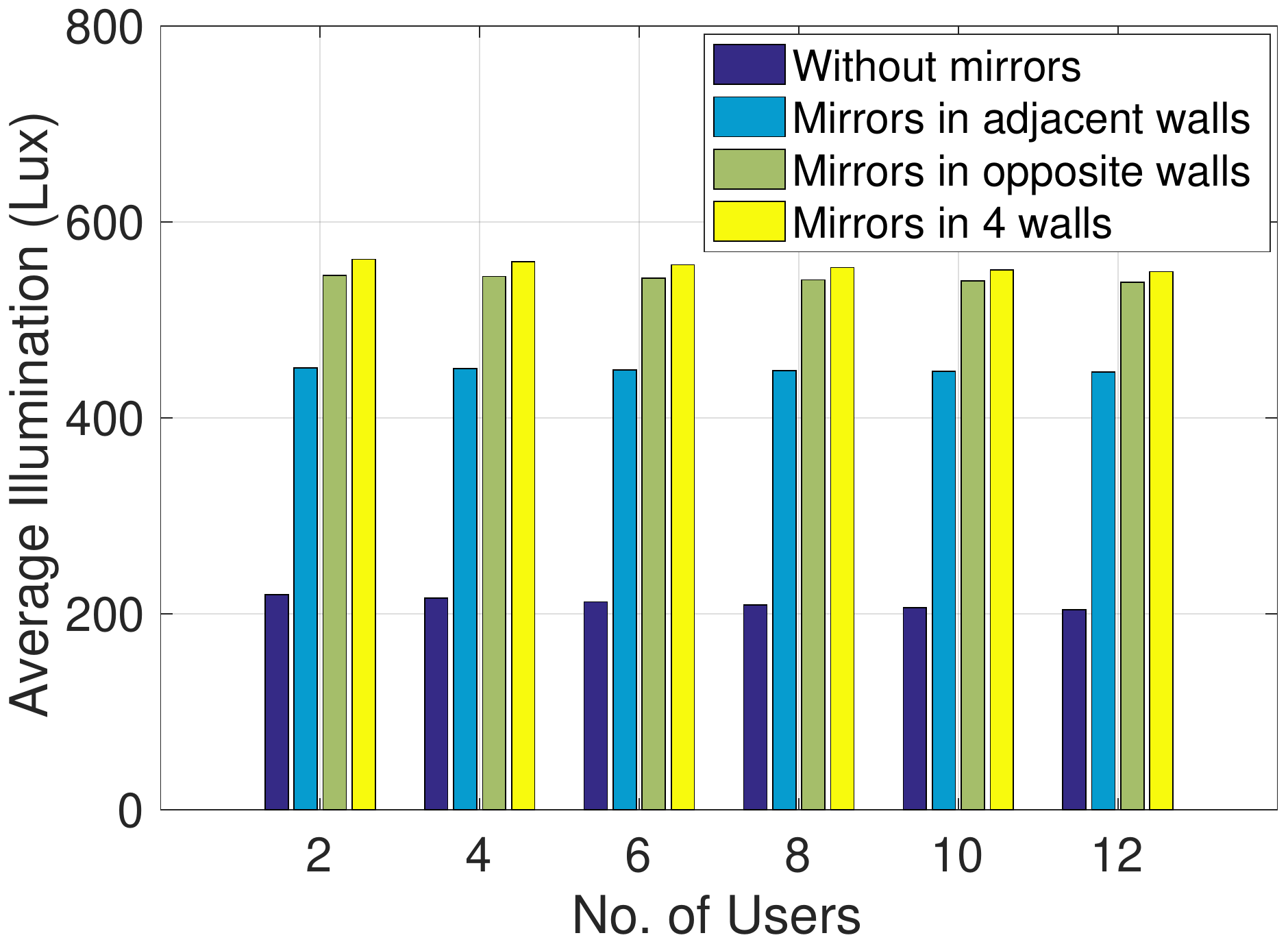}
 		\label{fig:ilum-nua}
 	}
 	\caption{Comparison between different mirror placement approaches for \textit{NUA} with $M=391$ and $\theta_d=30^o$.}
 	\label{fig:nua-comp}
 \end{figure*}

\begin{figure*}[t]
 	\centering
 	\subfigure[Minimum throughput vs. No. of users. ]{
 		\includegraphics[height=36mm,width=0.31\columnwidth] {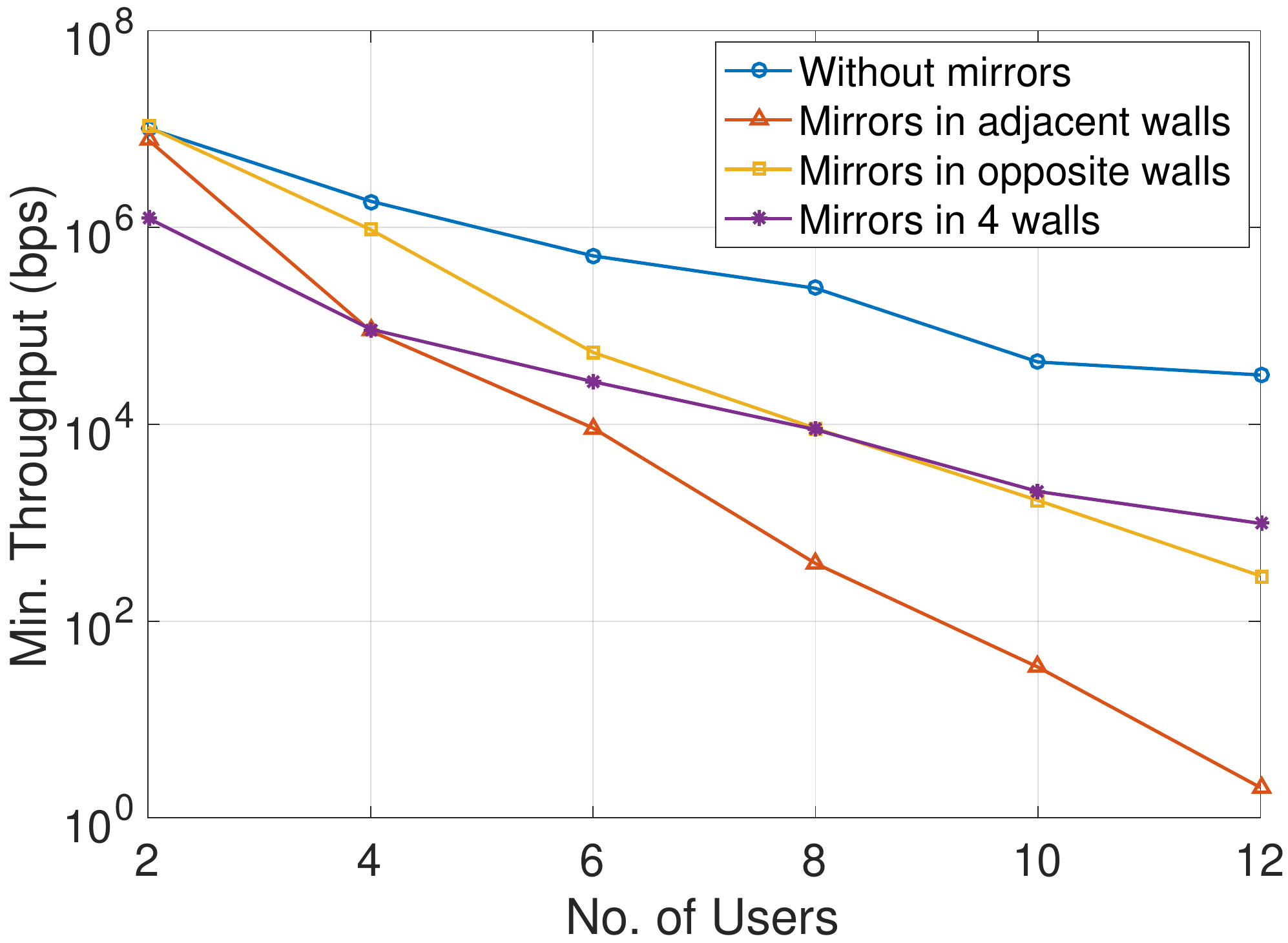}
 		\label{fig:mintp-ssa-user}
 	} 
 	\subfigure[Average throughput vs. No. of users.]{
 		\includegraphics[height=36mm,width=0.31\columnwidth] {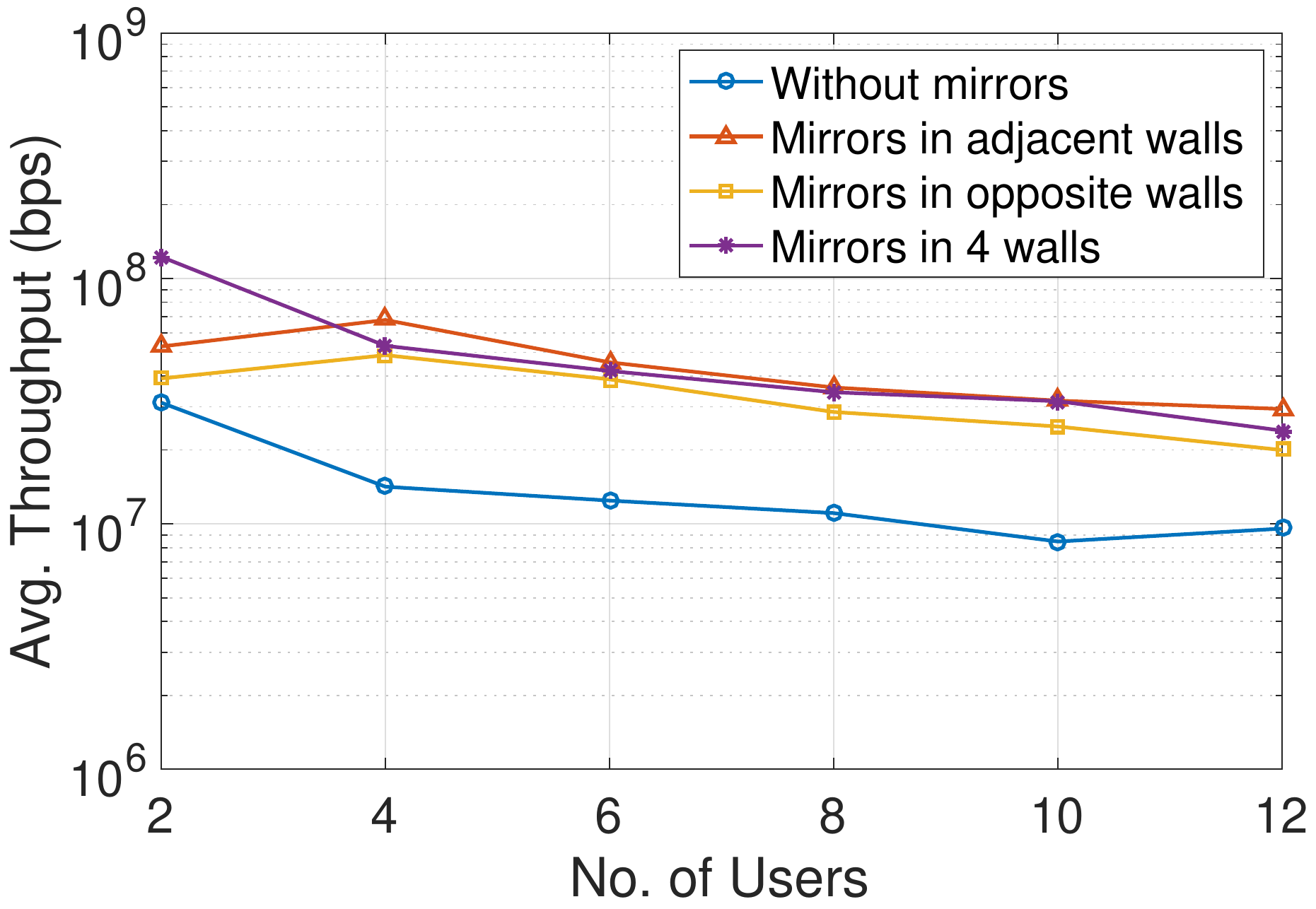}
 		\label{fig:avgtp-ssa-user}
 	} 	
 	\subfigure[Average illumination vs. No. of users.]{
 		\includegraphics[height=36mm,width=0.31\columnwidth] {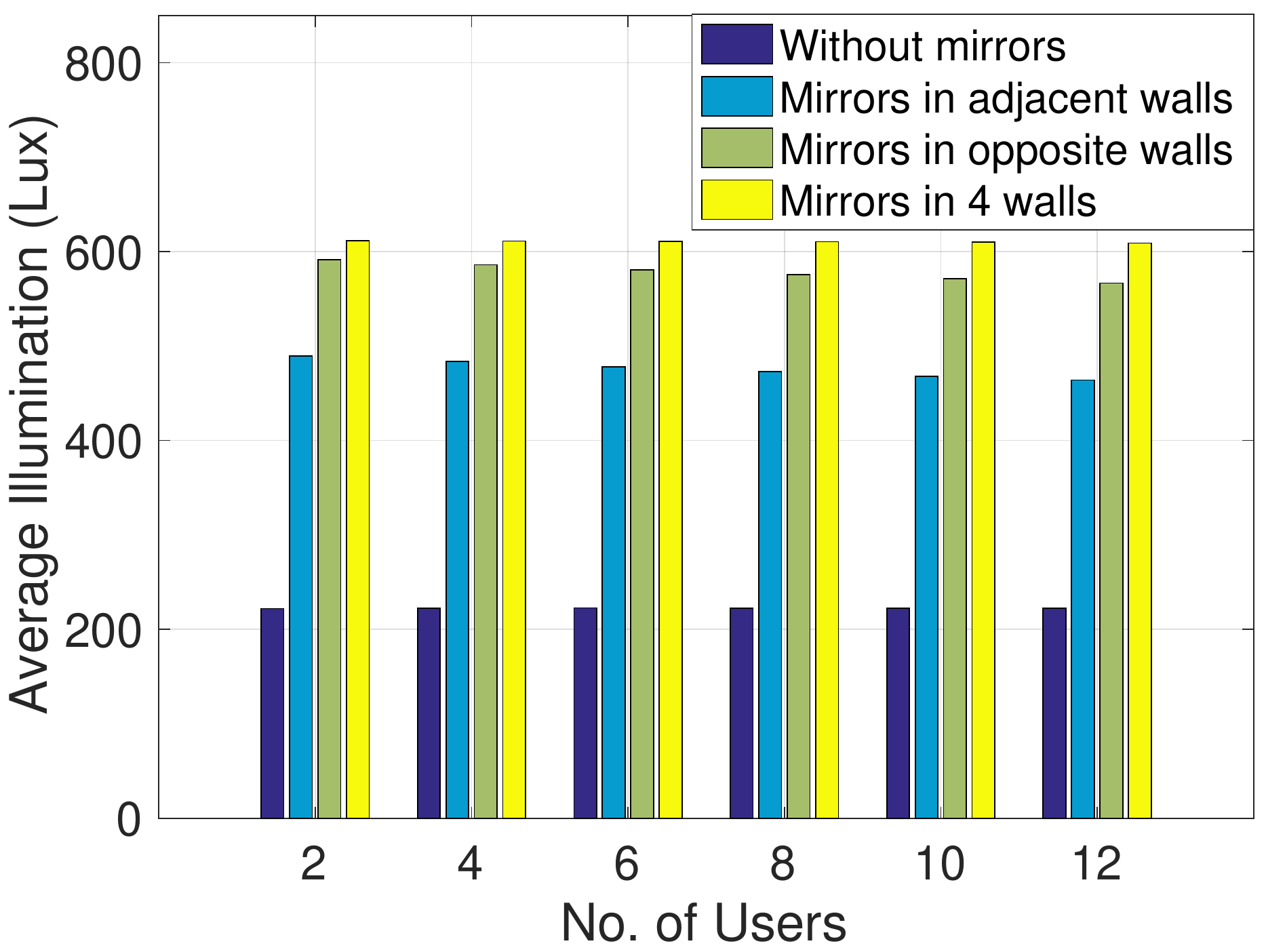}
 		\label{fig:ilum-ssa-user}
 	}
 	\caption{Comparison between different mirror placement approaches for \textit{SSA-User} with $M=391$ and $\theta_d=30^o$.}
 	\label{fig:ssa-user-comp}
 \end{figure*}

We compare the performance in terms of minimum throughput, average throughput and average illumination for the different mirror placement approaches. We also solve the design problem with no mirror placement at all to determine how much improvement is possible with mirror placement. The key findings are shown in Fig. \ref{fig:nua-comp} and \ref{fig:ssa-user-comp}. As the solution of the design problem is focused on optimizing the illumination, we see a noticeable three-fold improvement in average illumination level when mirror placement is applied, with the approach of placement in all walls having the most improvement aligning with our expectations. Although there is no significant improvement in minimum throughput after employing mirror placement, we can see up to four-fold increase in the average throughput values when comparing the four-wall case with the no mirror case. This fact indicates that placing mirror in all four walls can create more interference/noise for some users but at the same time they also provide better signal strength to most of the users resulting better average throughput. Adjacent walls and opposite walls cases also yield a solid improvement in both of average throughput and illumination level.

Overall, these comparisons provide valuable insights with regard to VLC network design. Since a much higher level of illumination 
can be obtained on average from the same number of LEDs, either of the three mirror placement approaches can be seen as beneficial. A trade-off between system performance and ease of mirror deployment can be observed among the approaches. Among the three approaches, placing mirror in all four walls is the one which can yield maximum performance improvement in terms of illumination. In terms of average throughput, all the three approaches demonstrate better performance compared to the no mirror approach. However, in terms of the ease of mirror deployment, placing mirrors in the opposite walls is arguably the preferable approach as the mirror placement pattern is such that most of the chosen mirror grids are connected to each other - in that case the group of connected mirrors can be placed as a larger mirror - making the deployment easier. In rest of the simulations, we use the four walls approach to obtain intermediate results from the design problem which are used to solve the communication problem.

 \begin{figure*}[t]
 	\centering
 	\subfigure[Minimum throughput vs. No. of users. ]{
 		\includegraphics[height=36mm,width=0.31\columnwidth] {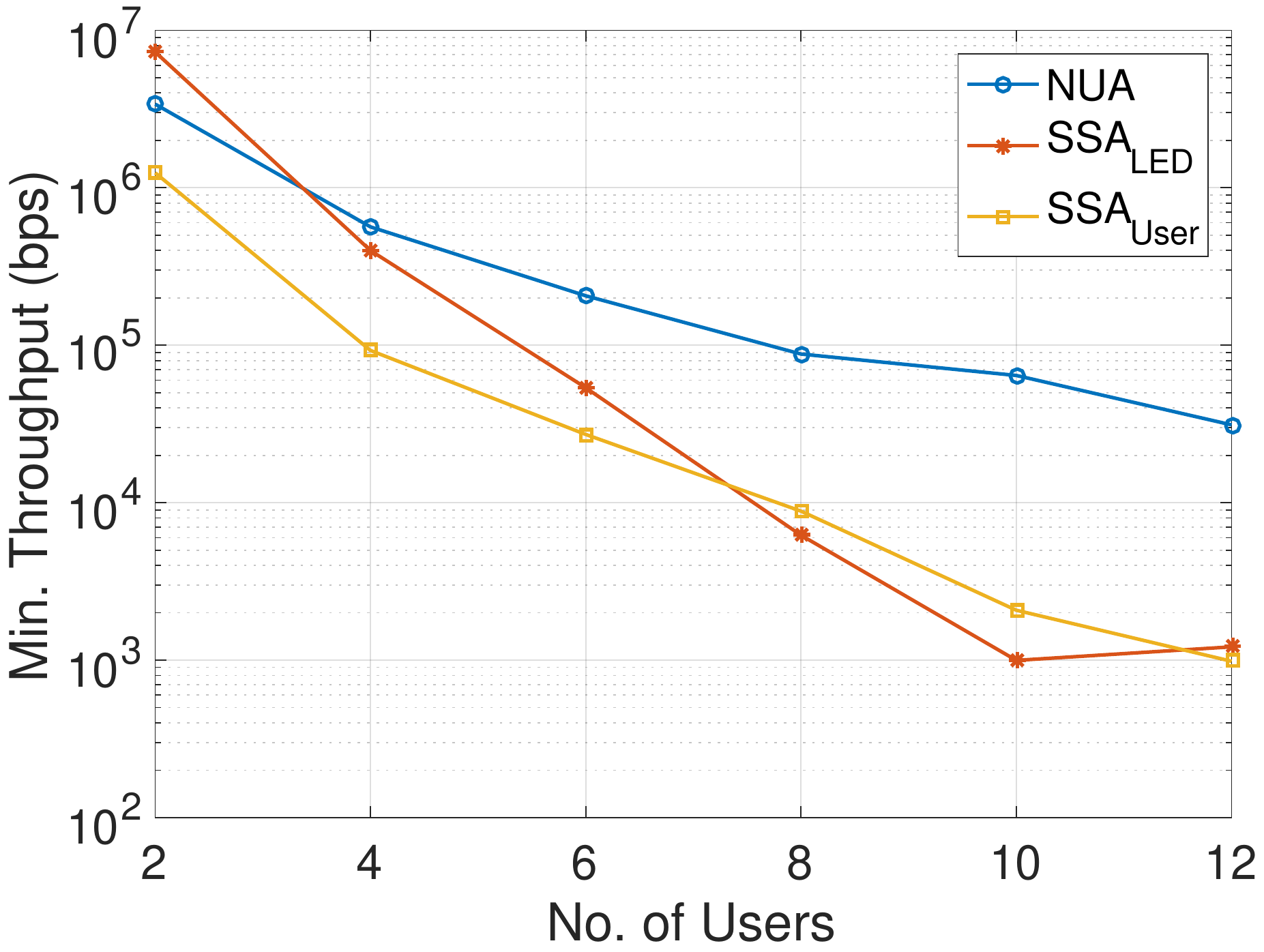}
 		\label{fig:comp_mintp}
 	} 
 	\subfigure[Average throughput vs. No. of users.]{
 		\includegraphics[height=36mm,width=0.31\columnwidth] {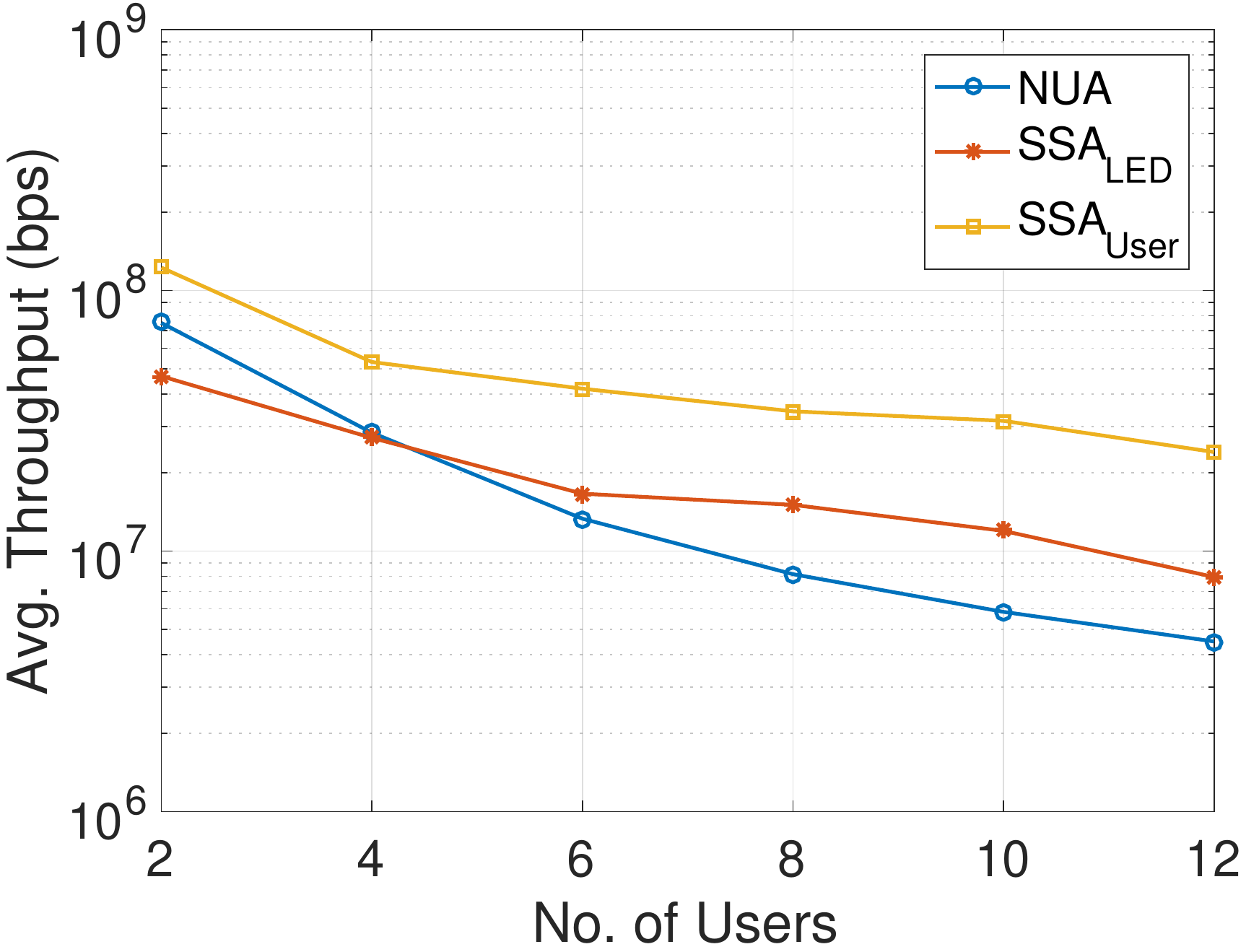}
 		\label{fig:comp_avgtp}
 	} 	
 	\subfigure[Average illumination vs. No. of users.]{
 		\includegraphics[height=36mm,width=0.31\columnwidth] {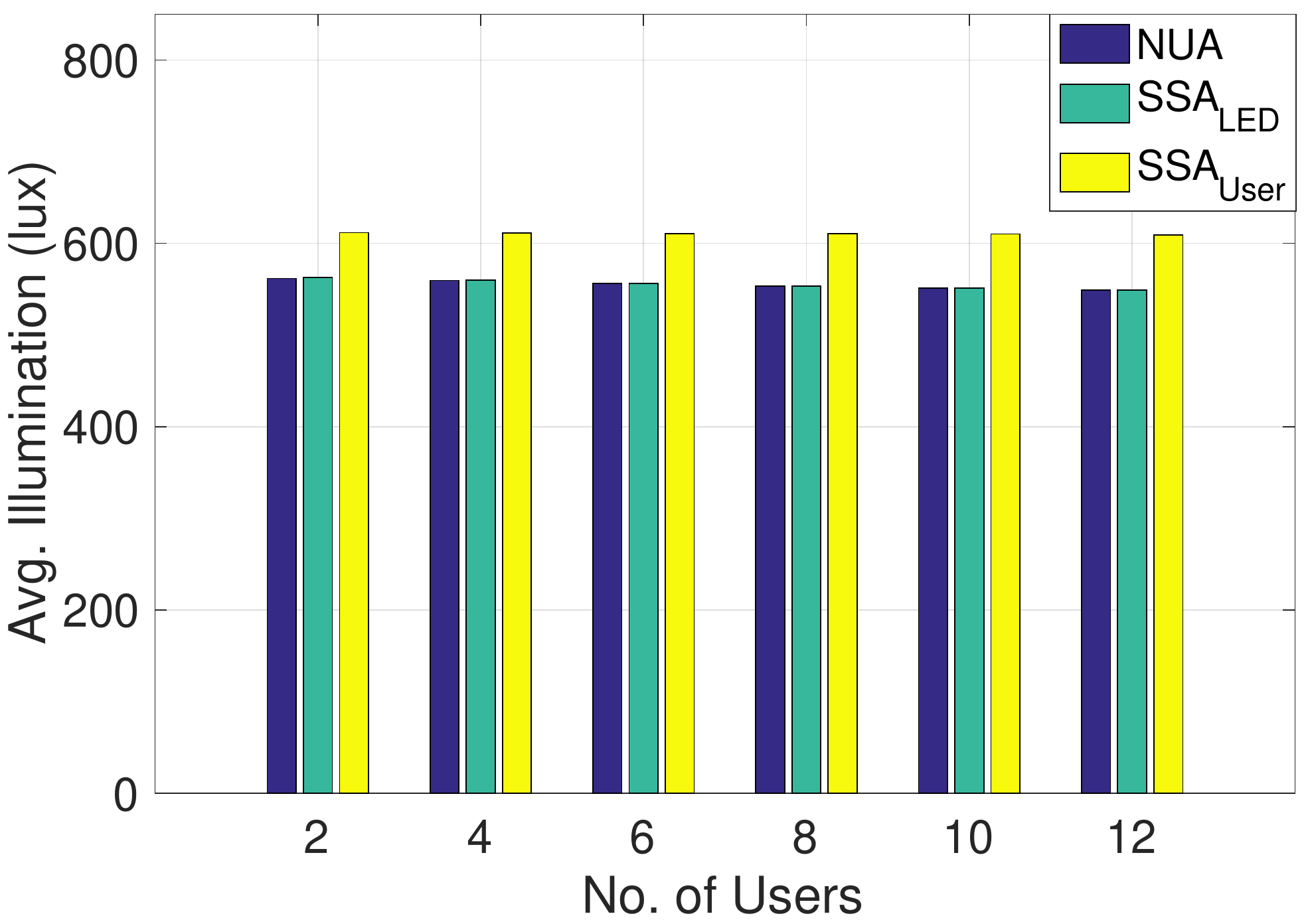}
 		\label{fig:comp_ilum}
 	}
 	\caption{Comparison between \textit{NUA}, \textit{SSA-LED} and \textit{SSA-User} with $M=391$ and $\theta_d=30^o$.}
 	\label{fig:comp}
 \end{figure*}


\subsection{Comparison between NUA, SSA-LED, and SSA-User}

We compare our proposed heuristic approaches for solving the communication problem, \textit{NUA}, \textit{SSA-LED} and \textit{SSA-User}. We compare both minimum and average throughput among the users with average illumination level across the sensing points from 2 to 12 users for these three approaches. As seen in Fig. \ref{fig:comp}, minimum  and average throughput decreases for all three approaches with increased number of users. For \textit{SSA-User}, minimum throughput has lower value compared to \textit{NUA} because this approach highly prioritizes the users chosen first for LED assignments, which largely affects the users chosen at the end. \textit{SSA-LED} also has lower values compared to \textit{NUA}, as the restriction of assignment for some of the LEDs leads some users to starvation. But the average throughput is better for \textit{SSA-LED} than \textit{NUA} as that restriction ensures a considerable decline in interference as well. \textit{SSA-User} does even better in terms of the average throughput as the high quality service provided to the users assigned first comprehensively overpowers the low quality service provided to the users in the end. In Fig. \ref{fig:comp_ilum}, we can see that the average illumination remains steady for all heuristics with increasing number of users since changes in LED source power allocation for these heuristics are not substantial enough to have any adverse effect on the overall illumination level.


 
 \begin{figure*}[t]
 	\centering
 	\subfigure[Minimum throughput vs. Divergence angle. ]{
 		\includegraphics[height=36mm,width=0.31\columnwidth] {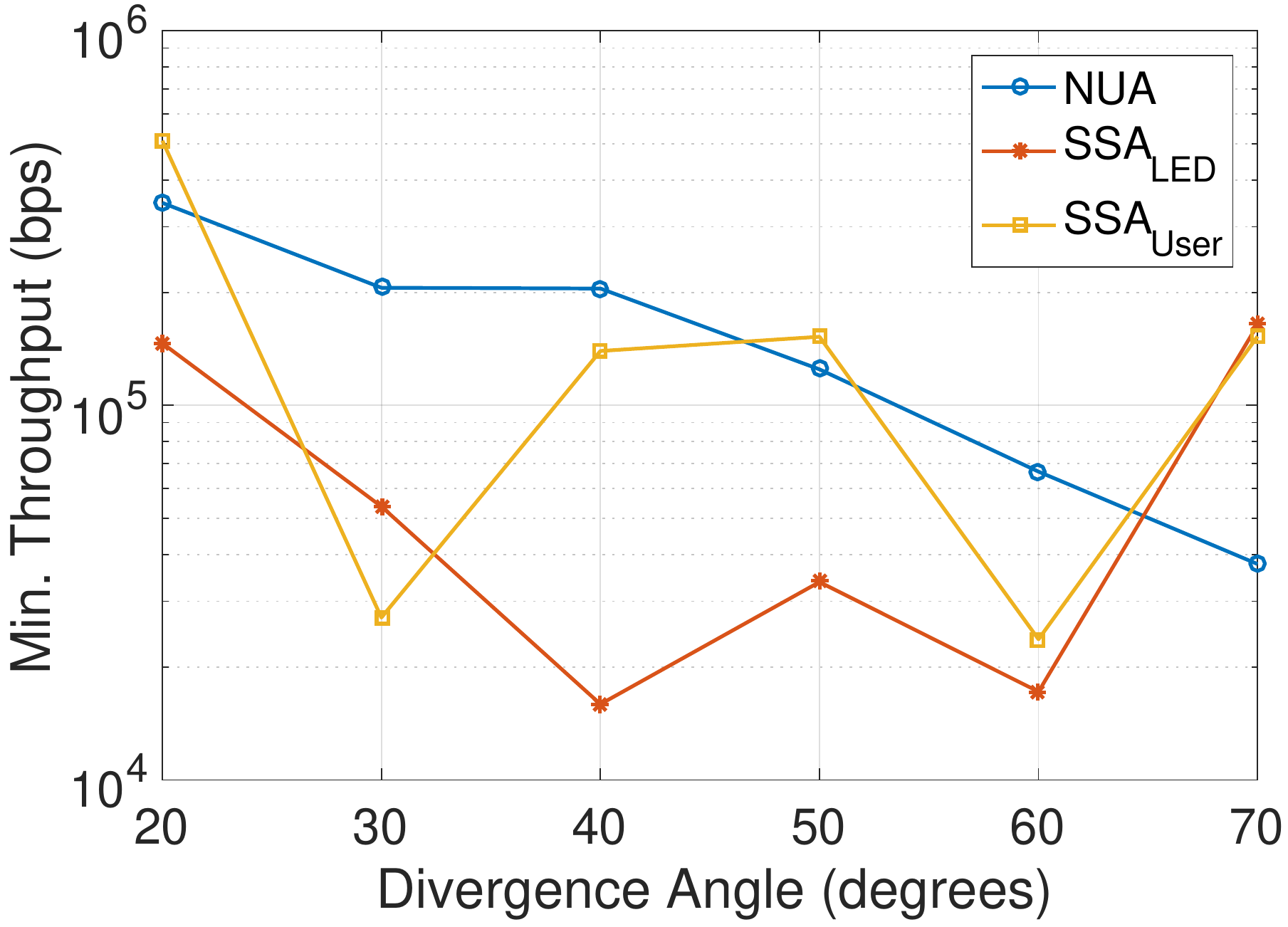}
 		\label{fig:ang_mintp}
 	} 
 	\subfigure[Average throughput vs. Divergence angle.]{
 		\includegraphics[height=36mm,width=0.31\columnwidth] {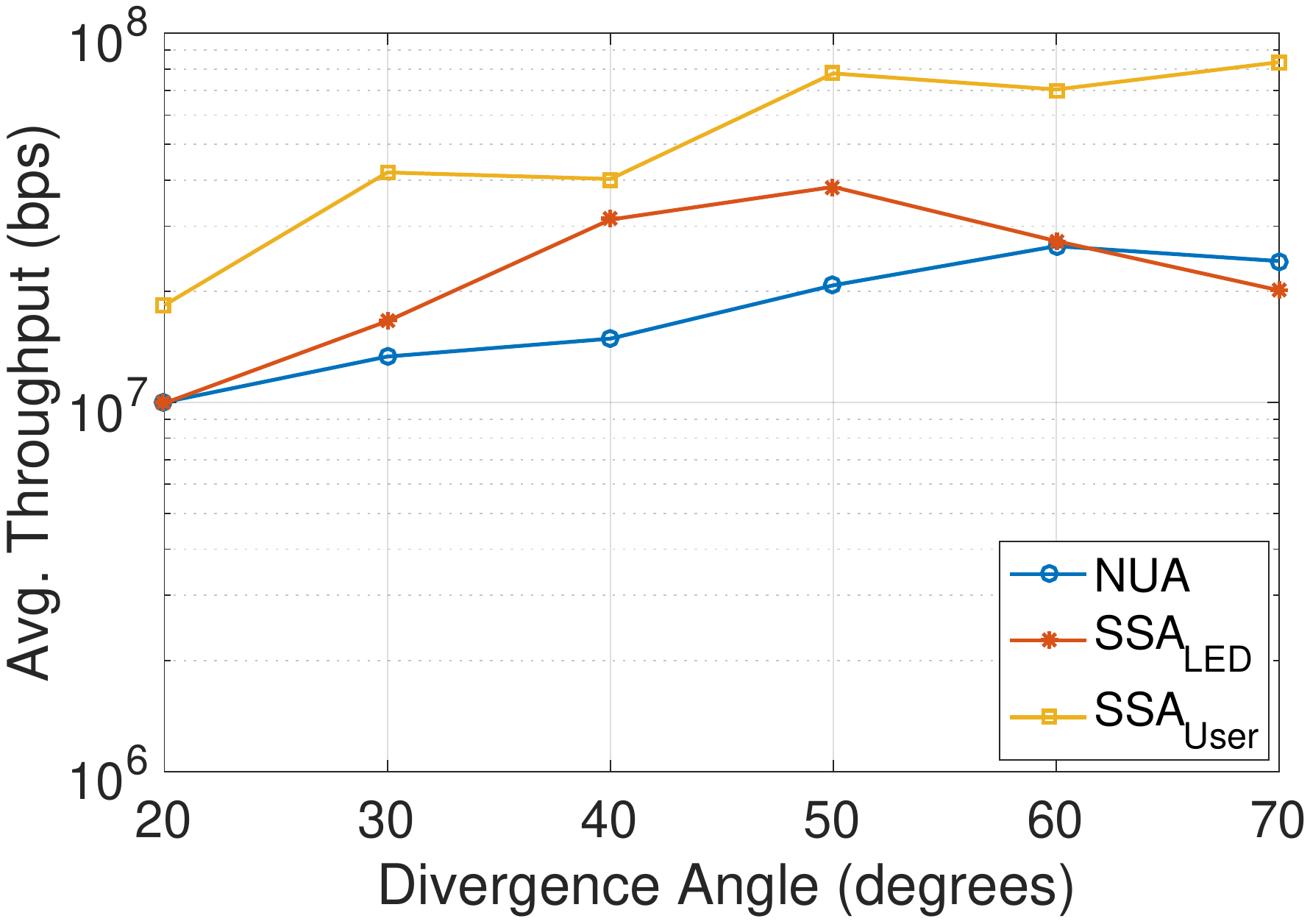}
 		\label{fig:ang_avgtp}
 	} 	
 	\subfigure[Average illumination vs. Divergence angle.]{
 		\includegraphics[height=36mm,width=0.31\columnwidth] {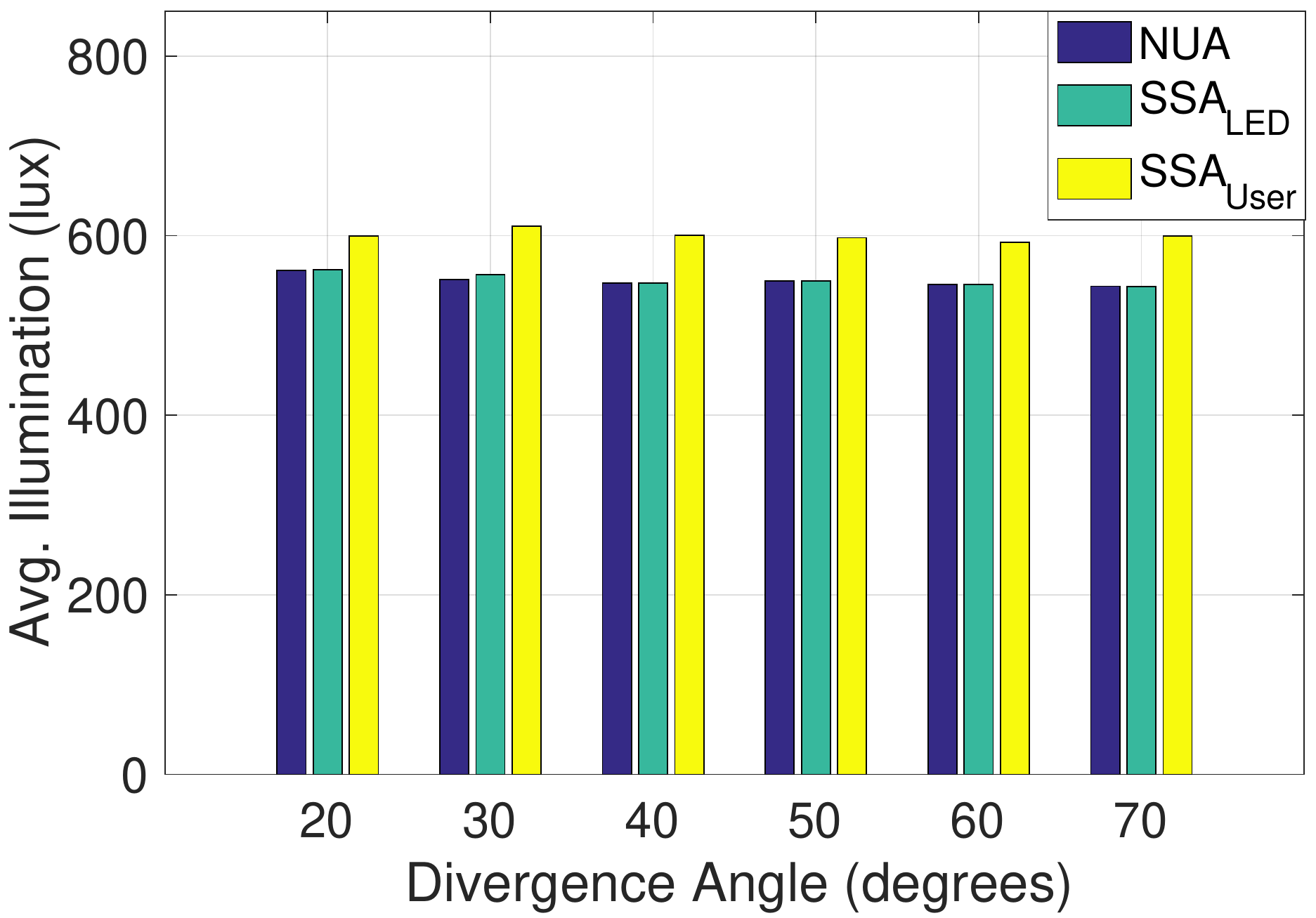}
 		\label{fig:ang_ilum}
 	}
 	\caption{Comparison between \textit{NUA}, \textit{SSA-LED} and \textit{SSA-User} for different divergence angle with $6$ users. $M=391$ and $R=40 cm$.}
 	\label{fig:ang-comp}
 \end{figure*}

\subsection{Effect of Divergence Angle}

For each of our proposed heuristics, we plot minimum and average throughput with average illumination in the room for different divergence angles when there are six users in total, shown in Fig. \ref{fig:ang-comp}. With larger divergence angles, the LED beams get larger, thus there are more chances of a particular user to  get coverage of an LED, but at the same time the chance of interference increases too. Changes in divergence angle does not seem have much impact on minimum throughput for \textit{SSA-LED} and \textit{SSA-User} because of the randomness in user locations. Looking at the average throughput, \textit{SSA-LED} and \textit{SSA-User} handle the interference better than \textit{NUA} for increasing divergence angles as shown in Fig. \ref{fig:ang_avgtp}. Average illumination level is not hampered much with either heuristic in this case as well.

\section{Summary and Future Work}
\label{sec:summary}

In this paper, we have explored a novel mirror employment approach for performance improvement for an MEMD VLC system. In order to maximize the average throughput and illumination, we have formulated an optimization problem to obtain optimum mirror placement, power allocation and LED-user association. As the problem is NP-complete, we have proposed a two-stage solution of the optimization problem. As a part of the solution of the first stage, we have introduced several mirror placement approaches and analyzed their performance. To deal with heavy computation complexity in the second stage with the communication problem, we have presented multiple heuristic approaches to solve it with a detailed analysis on their performance. We have shown that up to threefold increase in average illumination and fourfold increase in average throughput can be achieved using a mirror placement approach.

In light of our proposed mirror placement approach, many promising future works are possible in the realm of indoor MEMD VLC networks. For example, it is worth exploring different mirror sizes and shapes which might yield further improvement in throughput or illumination level. For the communication problem, other heuristic algorithms may be explored to attain results that are closer to the optimal solution. It would also be worth observing how the proposed algorithms and mirror placement approaches perform with different system parameters (e.g., room shape, shape of the bulb) and whether there are any relationship between them. In our work, we considered a single bulb in a standard size room. Designing a MEMD VLC system for a much larger indoor environment with multiple such bulb is definitely a challenging and promising direction.

	\bibliographystyle{IEEEtran}
	\bibliography{sifat-ref}

\end{document}